\documentclass[a4paper,fleqn,usenatbib]{mnras}

\usepackage{newtxtext,newtxmath}

\usepackage[T1]{fontenc}
\usepackage{ae,aecompl}

\usepackage{graphicx}	
\usepackage{amsmath}	
\usepackage{amssymb}	
\usepackage{longtable}
\usepackage{pdflscape}
\usepackage{listings} 
\usepackage{xcolor}
\definecolor{codegreen}{rgb}{0,0.6,0}
\definecolor{codegray}{rgb}{0.5,0.5,0.5}
\definecolor{codepurple}{rgb}{0.58,0,0.82}
\definecolor{backcolour}{rgb}{0.95,0.95,0.92}
\lstdefinestyle{mystyle}{
    backgroundcolor=\color{backcolour},   
    commentstyle=\color{codegreen},
    keywordstyle=\color{magenta},
    numberstyle=\tiny\color{codegray},
    stringstyle=\color{codepurple},
    basicstyle=\ttfamily\footnotesize,
    breakatwhitespace=false,         
    breaklines=true,                 
    captionpos=b,                    
    keepspaces=true,                 
    numbers=left,                    
    numbersep=5pt,                  
    showspaces=false,                
    showstringspaces=false,
    showtabs=false,                  
    tabsize=2
}
\title[deepSIP]{deepSIP: Linking Type Ia Supernova Spectra to Photometric Quantities with Deep Learning}

\author[B. E. Stahl et al.]{Benjamin E. Stahl,$^{1,2}$\thanks{E-mail: benjamin\_stahl@berkeley.edu}\thanks{Marc J. Staley Graduate Fellow.}
Jorge Mart\'inez-Palomera,$^{1}$
WeiKang Zheng,$^{1}$
\newauthor
Thomas de Jaeger,$^{1}$\thanks{Bengier Postdoctoral Fellow.}
Alexei V. Filippenko,$^{1,3}$
and Joshua S. Bloom$^{1,4}$
\\
$^{1}$Department of Astronomy, University of California, Berkeley, CA 94720-3411, USA\\
$^{2}$Department of Physics, University of California, Berkeley, CA 94720-7300, USA\\
$^{3}$Miller Senior Fellow, Miller Institute for Basic Research in Science, University of California, Berkeley, CA 94720, USA\\
$^{4}$Lawrence Berkeley National Laboratory, 1 Cyclotron Road, MS 50B-4206, Berkeley, CA 94720, USA
}

\date{Accepted XXX. Received YYY; in original form ZZZ}

\pubyear{2020}

\begin{document}
\label{firstpage}
\pagerange{\pageref{firstpage}--\pageref{lastpage}}
\maketitle

\begin{abstract}
We present {\tt deepSIP} (deep learning of Supernova Ia Parameters), a software package for measuring the phase and --- for the first time using deep learning --- the light-curve shape of a Type Ia supernova (SN~Ia) from an optical spectrum. At its core, {\tt deepSIP} consists of three convolutional neural networks trained on a substantial fraction of all publicly-available low-redshift SN~Ia optical spectra, onto which we have carefully coupled photometrically-derived quantities. We describe the accumulation of our spectroscopic and photometric datasets, the cuts taken to ensure quality, and our standardised technique for fitting light curves. These considerations yield a compilation of 2754 spectra with photometrically characterised phases and light-curve shapes. Though such a sample is significant in the SN community, it is small by deep-learning standards where networks routinely have millions or even billions of free parameters. We therefore introduce a data-augmentation strategy that meaningfully increases the size of the subset we allocate for training while prioritising model robustness and telescope agnosticism. We demonstrate the effectiveness of our models by deploying them on a sample unseen during training and hyperparameter selection, finding that Model~I identifies spectra that have a phase between $-10$ and 18\,d and light-curve shape, parameterised by $\Delta m_{15}$, between 0.85 and 1.55\,mag with an accuracy of 94.6\%. For those spectra that do fall within the aforementioned region in phase--$\Delta m_{15}$ space, Model~II predicts phases with a root-mean-square error (RMSE) of 1.00\,d and Model~III predicts $\Delta m_{15}$ values with an RMSE of 0.068\,mag.
\end{abstract}




\section{Introduction}
\label{sec:introduction}

The optical spectra of Type Ia supernovae (SNe~Ia) are rich with information \citep[for a review, see, e.g.,][]{Filippenko1997}. In addition to probing ejecta dynamics and chemical composition, spectral features have been found to encode the phase of a SN~Ia in its temporal evolution \citep[e.g.,][]{RiessSFA,FoleySFA,HowellSuperfit,snid,DASH}, and to a somewhat less quantitatively formalised extent, its peak luminosity \citep{Nugent1995,Si4000_lum_probe,Bailey2009,Blondin2011,bsnipIII,Zheng_empirical,kaepora}. The ability to extract the former (henceforth, the ``phase'') and the latter \citep[or something that correlates with it via a width-luminosity relation, such as $\Delta m_{15}$ or $\Delta$;][respectively]{Phillips1993,Riess1996} from optical spectra is of particular significance because both are conventionally derived from photometry. As the requisite light curves must consist of numerous individual observations conducted over at least several weeks, the ability to measure the aforementioned quantities from perhaps just a single observation (i.e., a spectrum) is of great value when allocating limited observing resources to optimise for specific science goals.

The SuperNova IDentification code \citep[{\tt SNID};][]{snid} has become the \emph{de facto}\footnote{As assessed from its prevalence in spectroscopic classifications issued by the Central Bureau of Electronic Telegrams (CBET) and in International Astronomical Union Circulars (IAUCs).} tool for classifying the type and phase of a SN from spectra, though alternatives do exist \citep[e.g., {\tt Superfit};][]{HowellSuperfit}. To determine the phase of a SN~Ia, such conventional approaches compare\footnote{{\tt SNID} uses cross-correlation \citep{Tonry1979} for comparison while {\tt Superfit} uses $\chi^2$ minimisation.} an input spectrum to a large database of spectra with known phases and then perform an aggregation of the phases from the best-matching templates. This approach has the advantage of being easy to understand (``SN X is most similar to SN Y at Z days relative to maximum brightness''), but it has the disadvantage of being inherently slow --- prediction time scales linearly with the number of template spectra in the database.

Machine learning \citep[ML; see ][for an overview of use cases in astronomy]{ML_astro} provides an interesting and fundamentally different approach to these tasks. In particular, phase and light-curve-shape determination can both be treated within the ``supervised learning'' paradigm, where a robust mapping between inputs and outputs is derived from a training set of input-output pairs. Subject to passing user-defined efficacy criteria when applied to a distinct testing set, the derived map can then be deployed to characterise new, unseen data. This approach leads to predictions that are fast (i.e., based on features themselves instead of comparisons against a large database) and therefore scalable. Accordingly, supervised ML has become increasingly prevalent in astronomical research campaigns \citep[e.g.,][]{BloomRF,MasciRF,GoldsteinRF,Wright2015,MillerRF,KimCNN,deepCR}.

Indeed, ML has proven to be a viable approach to photometric SN classification \citep[e.g.,][]{Richards-photoclassify,Moller2016,Lochner2016L,CM17,Narayan18,RAPID}, but only several studies thus far have applied such techniques to SN spectra. \citet{Sasdelli16} use \emph{unsupervised} ML techniques to explore the spectroscopic diversity of SNe~Ia, and find that much of the spectral variability, including that of the peculiar SN 1991bg-like \citep{91bg-Filippenko,91bg-Leibundgut} and SN 2002cx-like \citep[now known as the distinct ``SN~Iax'' class;][]{filippenko03,02cx-Li,FoleyIax} objects, can be parameterised by a carefully constructed five-dimensional space. As a much faster alternative to the aforementioned template-matching options (i.e., {\tt SNID}, {\tt Superfit}), \citet{DASH} have used a deep convolutional neural network \citep[CNN; see, e.g.,][]{deeplearning} to develop {\tt DASH}, a software package that classifies the type, phase, redshift, and host galaxy (but \emph{not} light-curve shape) of a supernova from optical spectra.

Motivated by this and the well-documented ability of CNNs to extract representative low-dimensional features from input signals, we formulate our approach as a set of three models, each of which utilises a similar CNN architecture to (Model I) determine if an input spectrum belongs to a SN~Ia within a specific domain in a space defined by phase and light-curve shape, (Model II) calculate the phase if it is within the domain, and (Model III) calculate a measure of the light-curve shape \citep[$\Delta m_{15}$;][]{SNooPy} if the same criterion is met. Although Model~II shares a common objective and architectural elements with {\tt DASH} (i.e., phase determination via a CNN architecture), we optimise specifically for SNe~Ia that fall within certain thresholds, treat the problem as one of \emph{regression} (not classification), and utilise dropout variational inference as a method by which to model uncertainties \citep{DVI,deepelement}. Moreover, our development of a CNN to predict the light-curve shape of a SN~Ia from its spectrum is novel.

We use the following sections to present the development of the aforementioned models. Section~\ref{sec:data} details the accumulation of our dataset, including how we process and prepare spectra for ingestion by our models. We outline our model architecture and discuss training and hyperparameter selection procedures in Section~\ref{sec:ML}, and we provide model-specific results in Section~\ref{sec:results}. Concluding remarks are then given in Section~\ref{sec:conclusion}.

\section{Data}
\label{sec:data}

\subsection{Spectra}
\label{ssec:spectra}

\begin{figure}
 \includegraphics[width=\columnwidth]{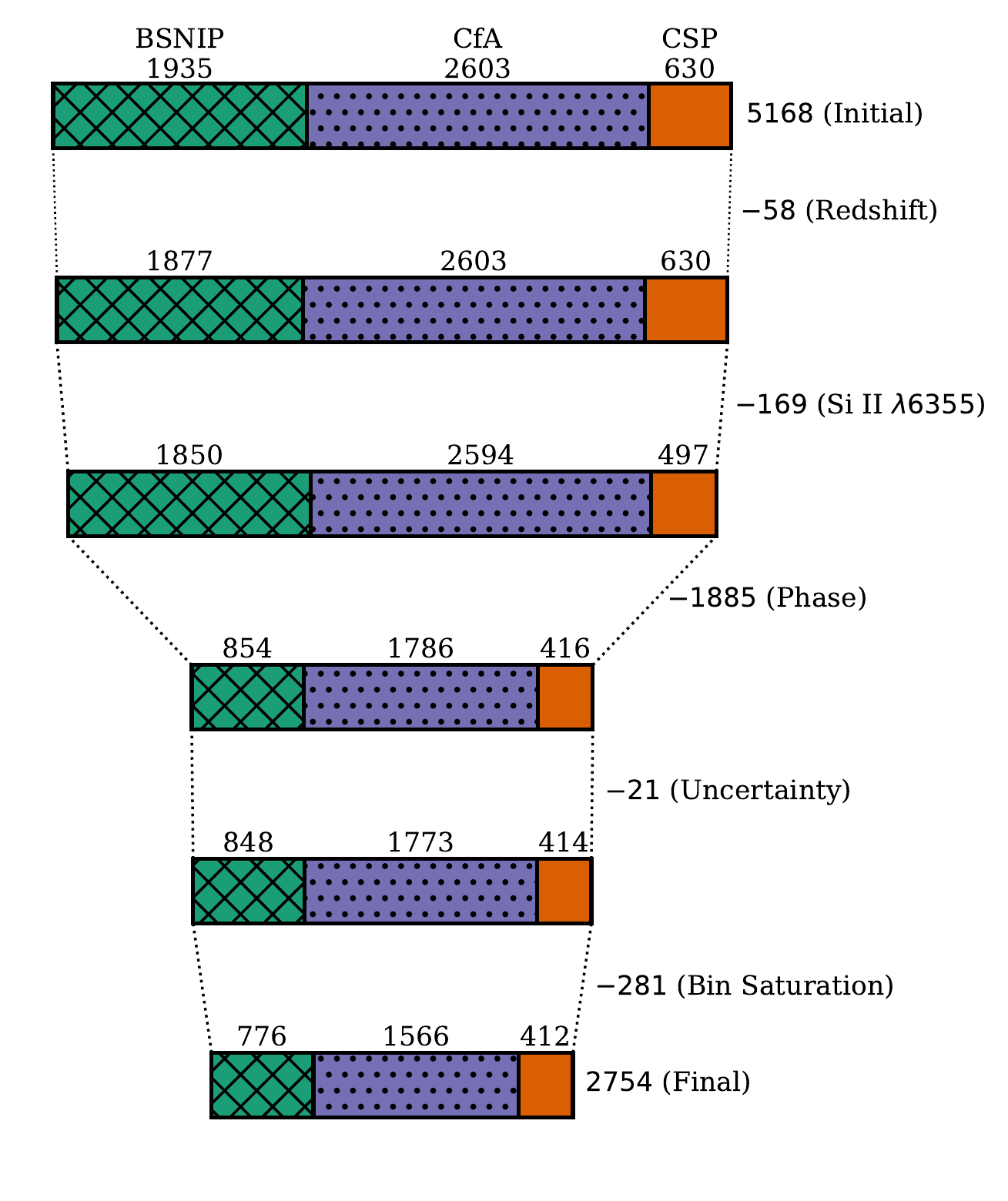}
 \caption{Full accounting of all cuts made in distilling our initial set of 5168 spectra down to the 2754 in our final compilation (1113 of which are within the ``domain'' defined in Section~\ref{ssec:final-compilation}). We delineate the source of each spectrum in the top row. By a wide margin, the lack of suitable photometric observations is responsible for the most severe cut (indicated with ``Phase'').\label{fig:cuts}}
\end{figure}

We source the spectra used herein from the three largest low-redshift SN~Ia spectral datasets currently in existence: the Berkeley SuperNova Ia Program \citep[BSNIP;][henceforth S12 and S20, respectively]{bsnipI,S20} sample with a total of 1935 spectra covering the period from 1989 through 2018 (see S12 for 1989--2008 and S20 for 2009--2018), the Harvard-Smithsonian Center for Astrophysics (CfA) sample with a total of 2603 spectra from observations spanning 1993--2008 \citep{BlondinB12}, and the Carnegie Supernova Program (CSP) sample with 630 spectra observed in the range 2004--2009 \citep{F13}. From this initial compilation of 5168 spectra, we perform two modest ``usability'' cuts that reduce our sample to 4941 (these cuts, in addition to those that are introduced below, are outlined in Figure~\ref{fig:cuts}). First, we drop the small fraction without a redshift listed in their associated publication, thereby yielding 5110 spectra, and second, we remove a further 169 that lack full coverage\footnote{We consider a SN~Ia spectrum to have full coverage of the \ion{Si}{ii} $\uplambda 6355$ feature if it has a minimum wavelength of less than 5750\,\AA\ and a maximum in excess of 6600\,\AA. These values represent the minimum and maximum extremes of the domains S20 use to search for the feature's blue and red endpoints, respectively.} of the \ion{Si}{ii} $\uplambda 6355$ feature that is ubiquitous in near-maximum-light SN~Ia spectra.

\begin{figure}
 \includegraphics[width=\columnwidth]{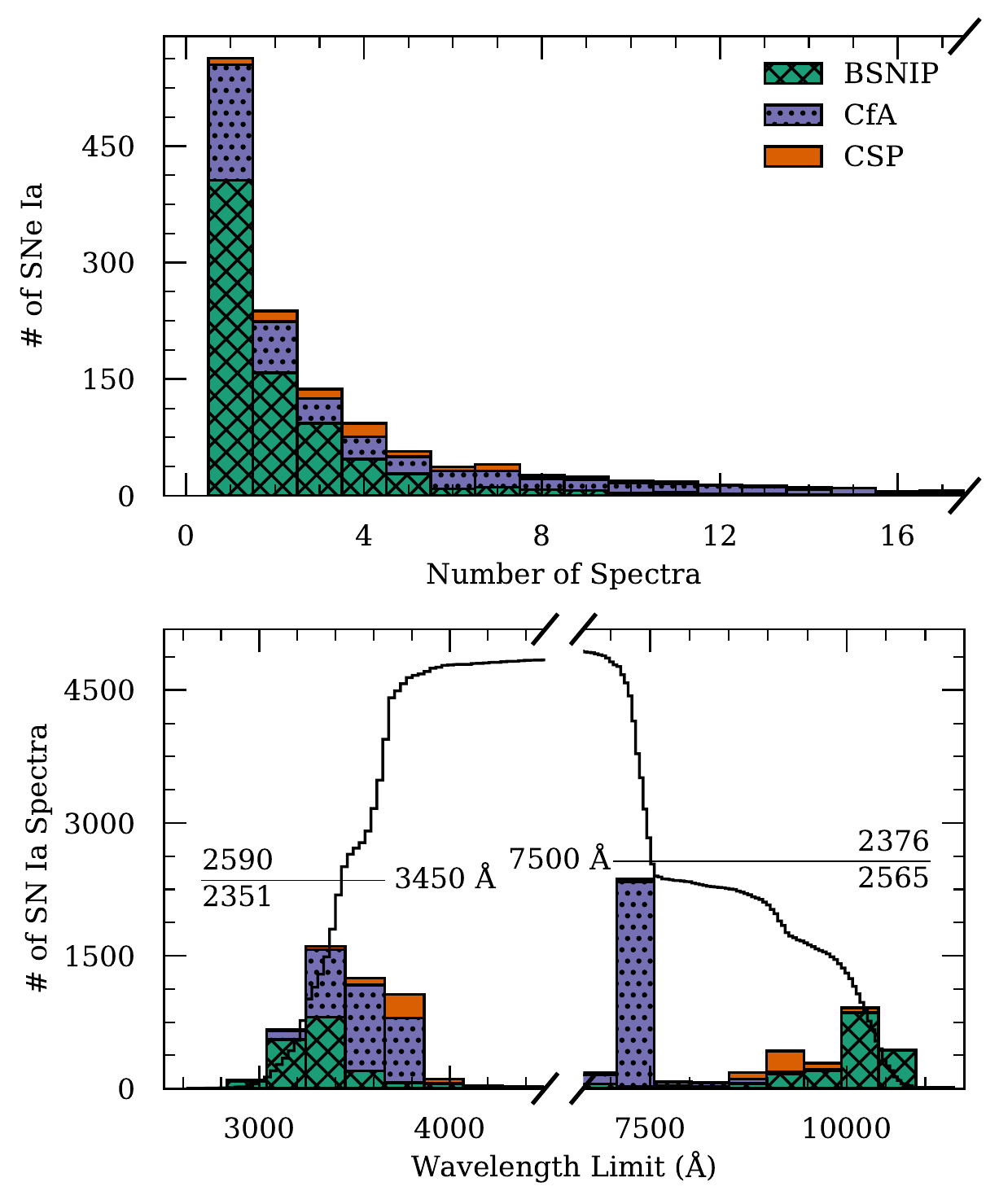}
 \caption{Stacked distributions of dataset parameters for our spectral compilation, distinguished by source. The top panel shows the number of spectroscopic observations per object (the tail extends to higher numbers of spectra, but is truncated for clarity) and the bottom panel displays the blue and red wavelength limits of the spectra. Overlaid on the bottom panel is the cumulative (inverse-cumulative) distribution of blue (red) wavelength limits, and the intersecting horizontal lines reflect the bounds defined in Section~\ref{sssec:pre-processing}. The number of spectra above and below each intersecting line are also labeled.\label{fig:nspec-wav-distr}}
\end{figure}

In addition to the high quality and sheer size of these datasets, the BSNIP and CfA sets were specifically selected for their complementarity --- whereas the observing strategy employed by the BSNIP is generally to prioritise the total number of SNe observed instead of the number of spectra \emph{per} SN, the CfA dataset covers fewer SNe but with higher cadence. This is clearly seen in the distribution of the number of spectra per SN in the top panel of Figure~\ref{fig:nspec-wav-distr}: the BSNIP sample spans many more SNe with several observations than does the CfA (or CSP) sample, but beyond $\sim 6$ spectra per object, the CfA sample wins out. Together, then, these datasets offer comprehensive coverage of the spectral diversity of SNe~Ia at both the individual and population levels.

We show the distribution of blue (red) wavelength limits for the spectra in our compilation in the lower panel of Figure~\ref{fig:nspec-wav-distr}. The superior red-wavelength coverage of the Kast double spectrograph on the 3\,m Shane telescope at Lick Observatory \citep[responsible for $\sim 79\%$ of the BSNIP sample;][]{Kast} to that of the FAST spectrograph on the 1.5\,m Tillinghast telescope at Whipple Observatory \citep[responsible for $\sim 94\%$ of the CfA sample;][]{FAST} is evident. The Lick spectra as well as most from CSP have good relative spectrophotometry owing to the slit being placed at the parallactic angle \citep{Filippenko1982}, but the continuum shapes of the FAST spectra may be inaccurate in some cases since the slit could not be rotated to arbitrary parallactic angles. Because any heterogeneities in the inputs to our models should reflect only physically significant information, we formulate our data preprocessing and augmentation procedures (see Sections~\ref{sssec:pre-processing} \&~\ref{sssec:augmentation}, respectively) to obscure as much source-specific information and contamination (e.g., wavelength limits, inaccurate continuum shapes) as possible.

\subsection{Light curves}
\label{ssec:light-curves}

As the purpose of this study is to identify and therefore derive, through supervised learning, certain photometrically-derived properties encoded in SN~Ia spectra, the aforementioned spectral compilation must be coupled to photometric observations (i.e., light curves), thereby allowing for the desired properties to be measured. To this end, we collect the requisite information from data releases by the same groups responsible for our compilation of spectra \citep[][henceforth G10 and S19 for the Berkeley sample, CfA1-3 for the CfA sample, and CSP3 for the CSP sample, respectively]{Ganeshalingam2010,S19,CfA1,CfA2,CfA3,CSP3}, as well as publish several new light curves (see Appendix~\ref{app:newlc}). We use the $E(B-V)$ model implemented within the {\tt SNooPy} package \citep[][see Appendix~\ref{app:snoopy} for additional details]{SNooPy} to fit the aforementioned light curves (except for those from S19 and CSP3, who have published fits using the same procedure along with their photometry), allowing us to measure the time of maximum \emph{B}-band brightness (and hence the phase\footnote{The phase of a spectrum is the time interval between when it is observed and when its SN reaches maximum \emph{B}-band brightness, as derived in Appendix~\ref{app:snoopy} and listed in Table~\ref{tab:snoopy-fits}, divided by a factor of $(1 + z)$ to correct for time dilation. The adopted redshift was listed in the original publication for that spectrum.}) and the decline-rate parameter\footnote{Our selected implementation of the SN~Ia width-luminosity relation uses a \emph{generalised} light-curve shape parameter, $\Delta m_{15}$, which is similar to --- but distinct from --- the more popular $\Delta m_{15}(B)$ used in the Phillips relation (i.e., the decline in magnitudes of a SN~Ia over the first 15\,d of its post-maximum $B$-band evolution). Indeed, the two may deviate randomly and systematically \citep[see Section 3.4.2 of][]{SNooPy}.}, $\Delta m_{15}$.

\subsection{Final Compilation}
\label{ssec:final-compilation}

\begin{figure}
 \includegraphics[width=\columnwidth]{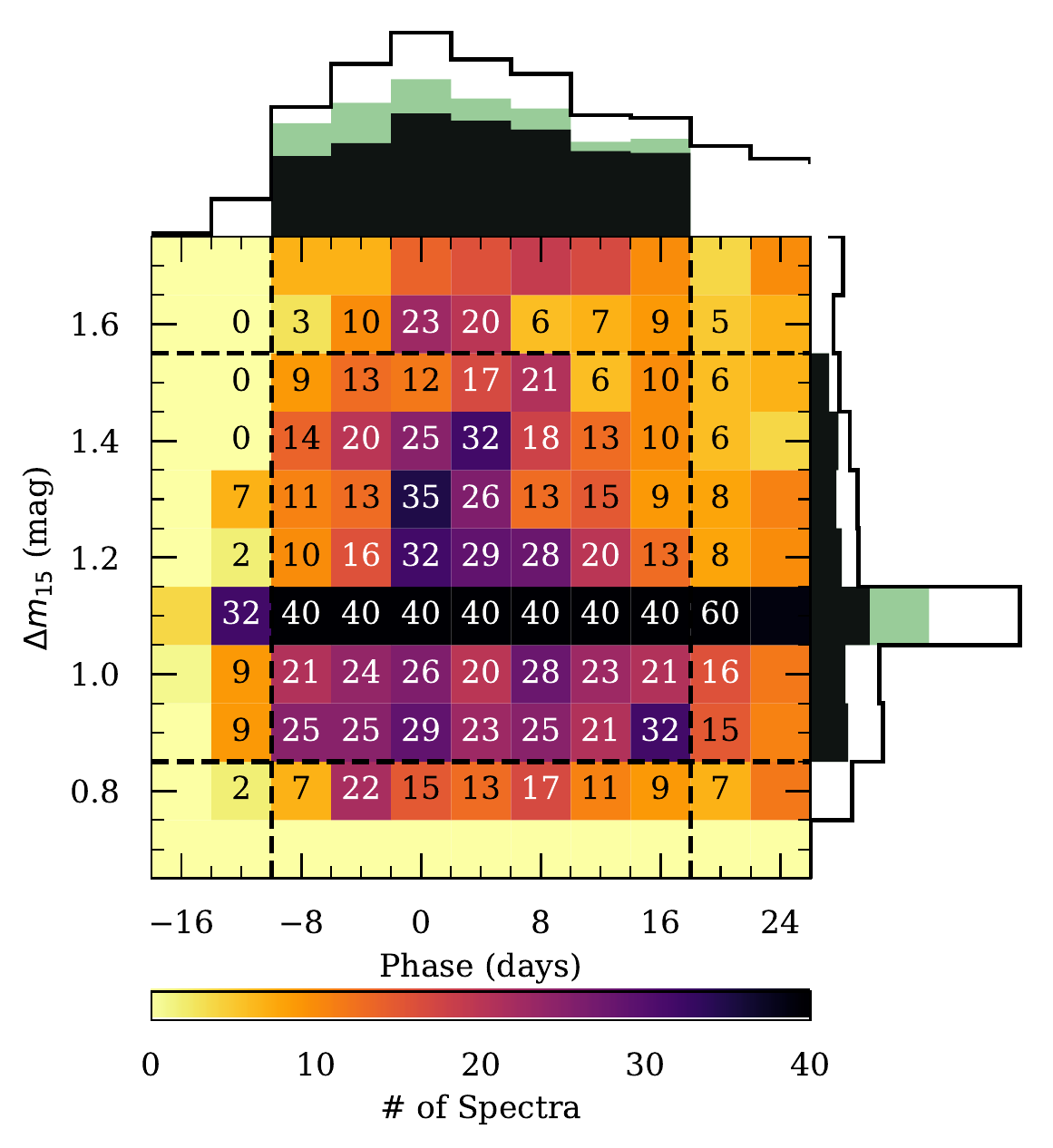}
 \caption{Distribution of $\Delta m_{15}$ and phase for the spectra in our compilation, with axes truncated to focus on the domain of interest. The empty-black one-dimensional projections reflect the full set of 3035 spectra, the green components consider only those 1394 spectra that are within the domain of interest, and the filled-black components show the same once spectra are removed to enforce the saturation criteria of 40 examples per bin (leaving 1113 in-domain spectra). The number of spectra falling within each in-domain bin and their immediate neighbors is labeled.\label{fig:dm15-phase-distr}}
\end{figure}

All told, 3056 spectra are linked to light curves with successful {\tt SNooPy} fits, but as shown in Figure~\ref{fig:cuts}, we remove 21 spectra having a phase uncertainty in excess of 1 day (d) and/or a $\Delta m_{15}$ uncertainty exceeding 0.1\,mag, thus yielding 3035 spectra. We visualise this compilation within the photometric parameter space of interest (namely, $\Delta m_{15}$ and phase) in Figure~\ref{fig:dm15-phase-distr}. Unsurprisingly, the densest coverage --- by a wide margin --- occurs for $\Delta m_{15} \approx 1.1$\,mag (reflecting that of a prototypical SN~Ia), but particularly impressive is the coverage within the region defined by $-10 \lesssim {\rm phase} \lesssim 18$\,d and $0.85 \lesssim \Delta m_{15} \lesssim 1.55$\,mag (albeit a bit sparse for the more rapidly declining objects within this region).

Motivated by this coverage, we impose the aforementioned region as a ``domain'' on our models in the following way: Model~I is tasked with classifying whether an input spectrum lies within its boundaries, while Model~II and Model~III determine the phase and $\Delta m_{15}$ (respectively) for spectra within this restricted domain. To mitigate the imbalance caused by the dominance of samples with $\Delta m_{15} \approx 1.1$\,mag, we enforce a ``saturation point'' of 40 samples for each in-domain bin in Figure~\ref{fig:dm15-phase-distr}. According to this policy, overly dense bins are brought into compliance by removing spectra with the largest $\Delta m_{15}$ uncertainties until only 40 remain. A total of 281 spectra are removed by this action, leaving 2754 examples (1113 of which are in-domain and thus relevant to Models II \& III) in what will henceforth be referred to as our final compilation (see the bottom row of Figure~\ref{fig:cuts}). Though this runs contrary to the common dogma that \emph{more data is always better}, we have found our choice to be empirically superior in this specific application.

\begin{figure*}
\centering
\includegraphics[width=\textwidth]{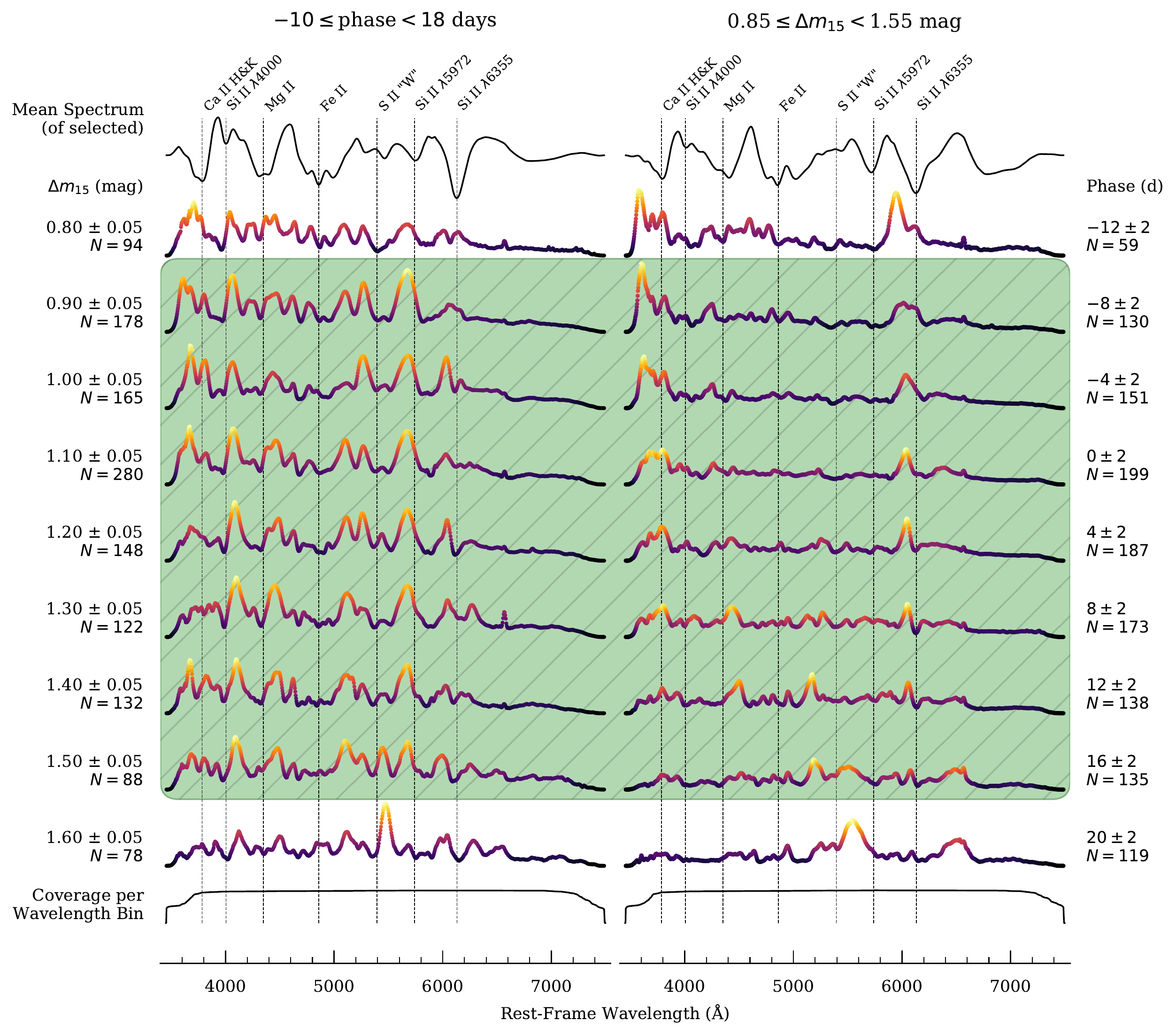}
\caption{Sequences of variance spectra progressing through equally spaced $\Delta m_{15}$ (left column) and phase (right column) bins. Each column begins with the selection criteria for the spectra in it and the mean spectrum of all those that are selected. Prominent spectral features are indicated. After advancing through the indicated variance sequence, the column terminates with the coverage per wavelength bin of the selected spectra. The same vertical scaling is applied to all variance spectra, but the colour map is normalised to each and is used to emphasise regions of significant variation. Our selected domain of interest is covered by the green region with diagonal hatching. All spectra used in generating the sequences have been preprocessed according to the specifications of Section~\ref{sssec:pre-processing}. The narrow spike that appears redward of the \ion{Si}{ii} $\uplambda 6355$ line in some variance spectra is due to nebular H$\alpha$ emission from the host galaxy.\label{fig:var-seq}}
\end{figure*}

A cursory inspection of Figure~\ref{fig:dm15-phase-distr} reveals that our coverage does not drop off significantly at larger phase and $\Delta m_{15}$ values than those which terminate our selected domain. It is therefore tempting to consider expanding the domain until such a drop is achieved (so as to make predictions over a wider swath of parameter space), but we choose not to do so for a myriad of reasons, the bulk of which are conveyed in the sequences of variance spectra presented in Figure~\ref{fig:var-seq}. If we assume that the spectral energy distribution (SED) of a SN~Ia is predominantly\footnote{We emphasise that ``predominantly'' does not mean ``exclusively'' --- other factors such as Galactic and host-galaxy extinction have an effect on an \emph{observed} SN~Ia SED;  our assumption is merely that those factors are of secondary significance to phase and light-curve shape, especially for spectra that have already been pre-processed in accordance with Section~\ref{sssec:pre-processing}.} determined by its phase and light-curve shape, then considering sequences of variance spectra --- whereby one of the aforementioned parameters is discretised into narrow bins and the variations within those bins are studied --- allows us to infer which regions in SN~Ia spectra vary the most at a given point in the sequence. If our assumption that the SED is largely a function of these two parameters holds, then such regions of large variation encode the most discriminating information about the nondiscretised parameter.

With this interpretation established, we note that for spectra with phases between $-10$\,d and 18\,d, the variance spectra in the left column of Figure~\ref{fig:var-seq} show notably similar structure for $\Delta m_{15}$ bins ranging from 0.90\,mag to 1.50\,mag. We interpret this as an indication that, at least within this range of $\Delta m_{15}$ values, phases between $-10$\,d and 18\,d are encoded by a common --- or ``slowly'' evolving --- set of features. A consequence of this is that a fairly simple convolutional neural network should be able to learn these features without much difficulty (we discuss our network architecture, including the way in which Figure~\ref{fig:var-seq} motives it, in more detail in Section~\ref{ssec:architecture}), and although a sophisticated network may well be able to learn ``when'' to weight certain features more heavily --- in addition to the features themselves --- we are content with the range of $\Delta m_{15}$ values afforded by our selected domain. Indeed, our coverage drops off sharply for lower $\Delta m_{15}$ and the more rapidly declining SNe~Ia in our dataset (i.e., those with $\Delta m_{15} \gtrsim 1.6$\,mag) are likely to be SN 1991bg-like objects which do not follow the Phillips relation (or its derivatives).

The aforementioned arguments do not perfectly carry over when we consider the phase-binned variance spectra for those SNe~Ia in our sample having $0.85 \leq \Delta m_{15} < 1.55$\,mag. Before maximum light, the blue wing of the \ion{Ca}{ii} H\&K feature exhibits the most variability and thus offers the best discrimination of $\Delta m_{15}$, but beyond peak, this variability fades and the dominant variation is observed in the blue wing of the \ion{Si}{ii} $\uplambda 6355$ feature. At phases $\gtrsim 10$\,d, this too begins to fade and variability is strongest at intermediate wavelengths, typically those in the vicinity of the \ion{S}{ii} ``W'' feature. It is beyond the scope of this study to speculate about --- or offer an explanation of --- the physical mechanism(s) that give rise to these observations, but we note that \citet{Nugent1995} identified these features in particular as a probe of SN~Ia luminosity, with the cause ascribed to temperature differences (and thus, to the total amount of $^{56}$Ni produced) between explosions. We do not pursue earlier phases owing to a paucity of data (indeed, Figure~\ref{fig:dm15-phase-distr} reveals that doing so would result in several empty bins), and while our compilation may well support an extension to later phases, we do not undertake such an addition here because Model~III would have to become very robust to evolving features. 

\subsubsection{Training, Validation, and Testing Sets}
\label{sssec:tvt}

In developing a neural network (or any supervised ML model), one typically divides the available data into three distinct subsets: a ``training'' set used to derive the decision path between features and outputs, a ``validation'' set to assess model performance during training and tune externally assigned hyperparameters, and finally, a completely separate ``testing'' set, which is used to probe the efficacy of the final model against unseen data, and \emph{not} used for either the optimisation of the network or the assignment of hyperparameters. In light of the small absolute size of our compilation (by modern ML standards), we intentionally set the validation and testing splits (10\% each) to be smaller than conventional allocations so as to keep our training set as large as possible.

We take a nuanced approach to ensure that our proportionally smaller validation and testing sets provide a realistic representation of our final compilation. Specifically, we sample according to a pseudostratified scheme for the 1113 in-domain spectra in our compilation, whereby we select random subsets of the appropriate size from each bin in Figure~\ref{fig:dm15-phase-distr}. In this way, the Model~II \& III training, validation, and testing sets have approximately the same binwise distribution. We impose a floor so that even bins with fewer than 10 total instances have at least one sample for each of the validation and testing sets. As a result, the actual validation and testing ratios are elevated slightly higher than the targeted 10\%. The Model~I sets are generated by randomly sampling all out-of-domain spectra at the prescribed ratios and then adding them to the pseudostratified in-domain sets. This ensures that all spectra in the Model~II/III sets are just subsets of the corresponding Model~I sets. Therefore, we can holistically assess {\tt deepSIP} via the Model~I testing set without fear of Model~II or III inadvertently being asked to characterise spectra that occur in their training or validation sets.

\subsubsection{Preprocessing}
\label{sssec:pre-processing}

\begin{figure}
 \includegraphics[width=\columnwidth]{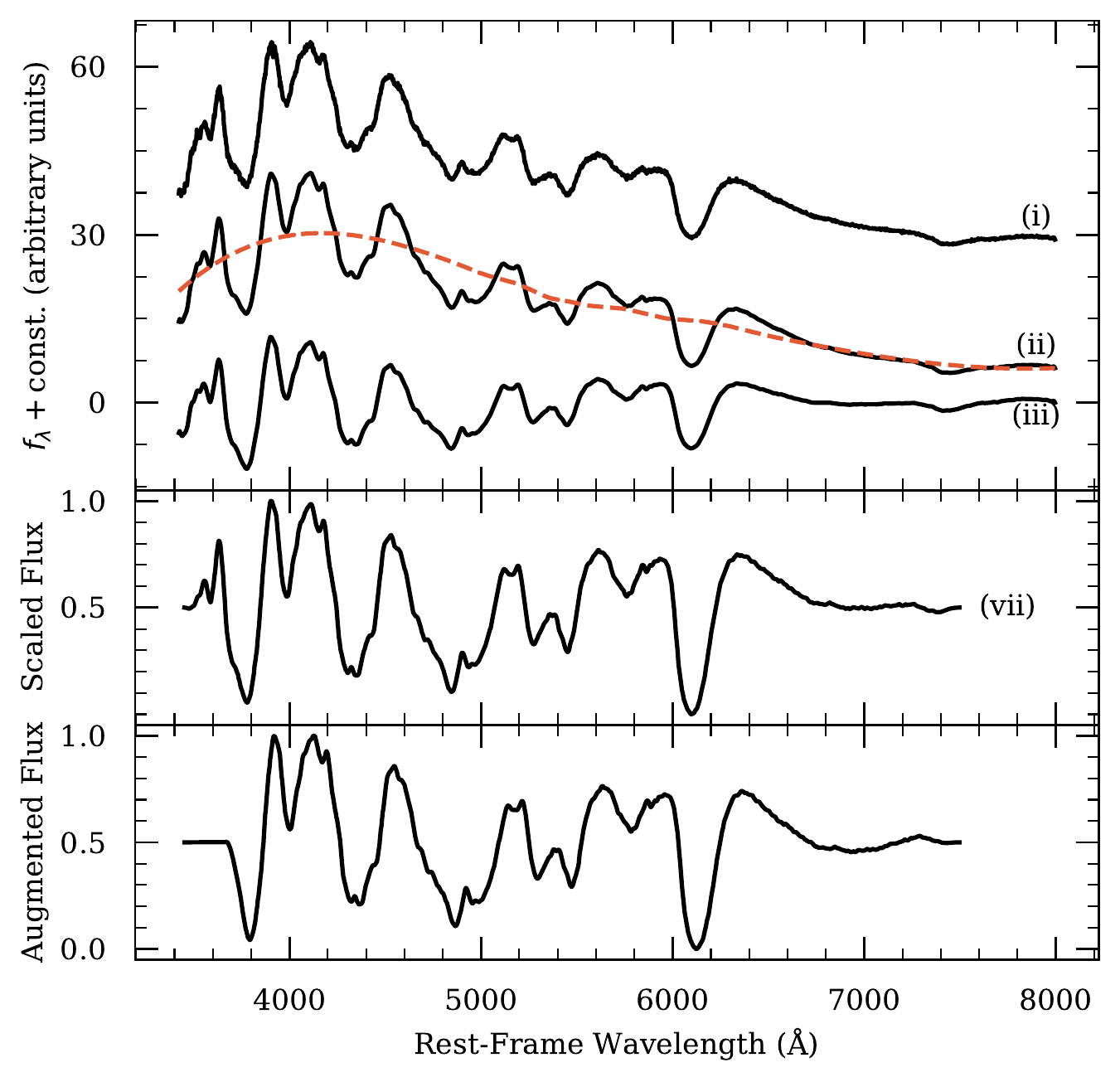}
 \caption{Snapshots showing various stages of our preprocessing routine as applied to a spectrum of SN~2016coj at $+ 1.3$\,d. The numerals indicate the last preprocessing step to have been performed on the plotted spectrum, and the dashed orange line illustrates the fitted (and subsequently removed) pseudocontinuum. The bottom panel shows an example of the spectrum after data augmentation steps have been applied.\label{fig:preprocessing}}
\end{figure}

Our models should be sensitive only to the physical characteristics encoded in the spectra they are trained on, not to any peculiarities relating to how the spectra were collected or reduced. Furthermore, it is imperative that each spectrum is processed in a carefully controlled and systematic way to avoid inadvertent biases. We therefore perform the following preprocessing steps to homogenise input spectra prior to ingestion by our models.
\begin{enumerate}
	\item Each spectrum is de-redshifted --- that is, the redshift is removed. This step is skipped for augmented spectra (see Section~\ref{sssec:augmentation}) which are already in (or near) the rest frame.
	\item Each spectrum is smoothed using a Savitzky-Golay filter \citep{SGfilter} with a window equivalent to 100\,\AA, though the window is varied for augmented spectra.
	\item The pseudocontinuum is modeled by again smoothing the spectrum, but with a much wider window of 3000\,\AA~(unless this exceeds the range of the spectrum, in which case we use a dynamically determined value corresponding to $\sim 70\%$ of the available wavelength range). We then subtract it from the spectrum.
	\item The spectrum is binned onto a log-wavelength scale consisting of 1024 points between 3450\,\AA\ and 7500\,\AA. As shown in Figure~\ref{fig:nspec-wav-distr}, these endpoints are such that $\sim 50$\% of our global compilation (i.e., including those \emph{without} phase information) have additional spectral information either below 3450\,\AA\ or above 7500\,\AA\ that is disregarded. This painful step of throwing away potentially useful information is necessary to avoid inducing significant biases between our data sources. If a spectrum does not have signal all the way to the blue or red ends of this range, we set it to zero in the missing end(s). In addition to ensuring that all spectra are represented by vectors of the same length, this transformation has the useful consequence that redshifting corresponds to a linear translation (see Section~\ref{sssec:augmentation} for more details).
	\item We scale the signal so that it has a range of unity and then translate it such that it has a mean of zero.
	\item The first and last 5\% of the signal in the spectrum is tapered using a Cosine Bell (i.e., a Hanning window) so that it smoothly goes to zero at the ends.
	\item Finally, we add 0.5 to the signal so that it is positive everywhere. Henceforth, we refer to this quantity as ``scaled flux.''
\end{enumerate}
We show an example of the intermediate stages and final result derived from our preprocessing procedure in Figure~\ref{fig:preprocessing}.

\subsubsection{Augmentation}
\label{sssec:augmentation}

Though we have taken care to assemble a significant fraction of \emph{all} publicly available low-redshift SN~Ia optical spectra currently in existence, our final compilation is still rather small by modern standards in deep learning (especially for the domain-restricted subset that is relevant for Models~II and III). For this reason, we formulate a data-augmentation strategy \citep[i.e., a method for extending our training set beyond its limited size while preserving its characteristics; e.g.,][]{2015MNRAS.450.1441D,2017ApJ...836...97C,2018AJ....156..186M,2019AJ....158..257B} that generates a training set of substantially increased size. To accomplish this, we randomly sample data from the Model~I (II/III) training set, with replacement, until we have a collection whose size, when combined with the non-augmented training set, equals 5000 examples (a $\sim 4$-to-1 ratio of augmented to original training samples for the Model II/III set). After obtaining samples according to this prescription, we transform each sampled spectrum using the following operations.

\begin{enumerate}
	\item \textbf{Redshifting:} As noted in Section~\ref{sssec:pre-processing}, we remove the redshift from all spectra that are fed into our models. However, we expect our models to be robust to small redshift errors that propagate into the rest-wavelength transformation. To this end, we perturb the rest wavelength array of each sampled spectrum by a multiplicative factor of $(1 + \delta z)$, where $\delta z$ is drawn from a uniform distribution, $\delta z \sim \mathcal{U}(-0.004, 0.004)$, motivated by the mean uncertainty in the {\tt SNID}-derived redshifts reported by S20 for their dataset. Coupled with log-binning (which converts redshifting/de-redshifting into a linear offset; see Section~\ref{sssec:pre-processing}), this allows us to reinforce and exploit the invariance to small translations that CNN architectures exhibit \citep{deeplearning}.
	\item \textbf{Noise:} To encourage our models to be robust to variations in the signal-to-noise ratios of input spectra, we vary the degree of smoothing applied to each sampled spectrum during our preprocessing procedures (see Section~\ref{fig:preprocessing}). We do this by randomly selecting the smoothing window, $w \sim \mathcal{U}(50, 150)$\,\AA, with these bounds chosen to be roughly consistent with the range of wavelength-space extents of the high-variance regions identified in Figure~\ref{fig:var-seq}.
	\item \textbf{Trimming:} We expect our models to be insensitive to any information about the observing apparatus and configuration that might be encoded in a spectrum. For example, the median phase of the BSNIP-collected spectra in our compilation is $\sim 18$\,d, while for the CfA-collected spectra it is $\sim 9$\,d, but our models should \emph{not} form a decision path that preferentially associates spectra having extensive red-wavelength coverage (namely, the BSNIP spectra) with later phases --- this correlation is purely a consequence of exogenous biases in our compilation. Therefore, in addition to the preprocessing steps outlined in Section~\ref{sssec:pre-processing}, we remove a random proportion, $f$, from the blue and red ends of each sampled spectrum, where $f \sim \mathcal{U}(0, 0.1)$, with the upper bound chosen so as to maintain full coverage of the characteristic \ion{Si}{ii} $\uplambda 6355$ feature that we require for spectra in our compilation to possess.
\end{enumerate}
We show an example of the results of the aforementioned augmentation procedures in the bottom panel of Figure~\ref{fig:preprocessing}.

\section{Models}
\label{sec:ML}

As noted, we have constructed three models to ultimately determine the phase and light-curve shape parameter, $\Delta m_{15}$, of a SN~Ia from an optical spectrum. The first model determines if the input is from a SN~Ia with a phase of $-10 \leq {\rm phase} < 18$\,d that has a light-curve-shape parameter of $0.85 \leq \Delta m_{15} < 1.55$\,mag. The second model determines the phase, and the third, $\Delta m_{15}$, both only within the domain for which Model~I discerns. We formulate the first model as a binary classification problem --- either a spectrum belongs to a SN~Ia subject to the aforementioned photometric restrictions, or it does not. The remaining models can be construed as a regression problem, where a continuous quantity (e.g., phase or $\Delta m_{15}$) is to be predicted. Despite their differing applications, each model uses a similar neural network architecture, and much of the work flow of training and evaluating them is common. We therefore devote the following subsections to discussing the underlying architecture employed in our models and the common aspects of our work flow. We present model-specific results in Section~\ref{sec:results}.

\subsection{Architecture}
\label{ssec:architecture}

At the heart of our models is a deep (i.e., multilayer) CNN, but whereas the prototypical use-case is two-dimensional (2D) --- deep 2D CNNs have a storied history in image classification \citep[e.g.,][]{LeCun90,LeCun98,GoogLeNet} and even \citet{DASH} resorted to tiling 1D SN spectra into 2D ``images'' to formulate their problem as one of image identification --- we follow the inherent dimensionality of our data by using a 1D implementation \citep[for a summary of 1D CNNs, see][]{1DConv}. In addition to being more conceptually compatible with our application, our use of a 1D CNN gives us much more control over the degree to which nonlocal features are aggregated though pooling operations.

\begin{figure*}
 \includegraphics[width=\textwidth]{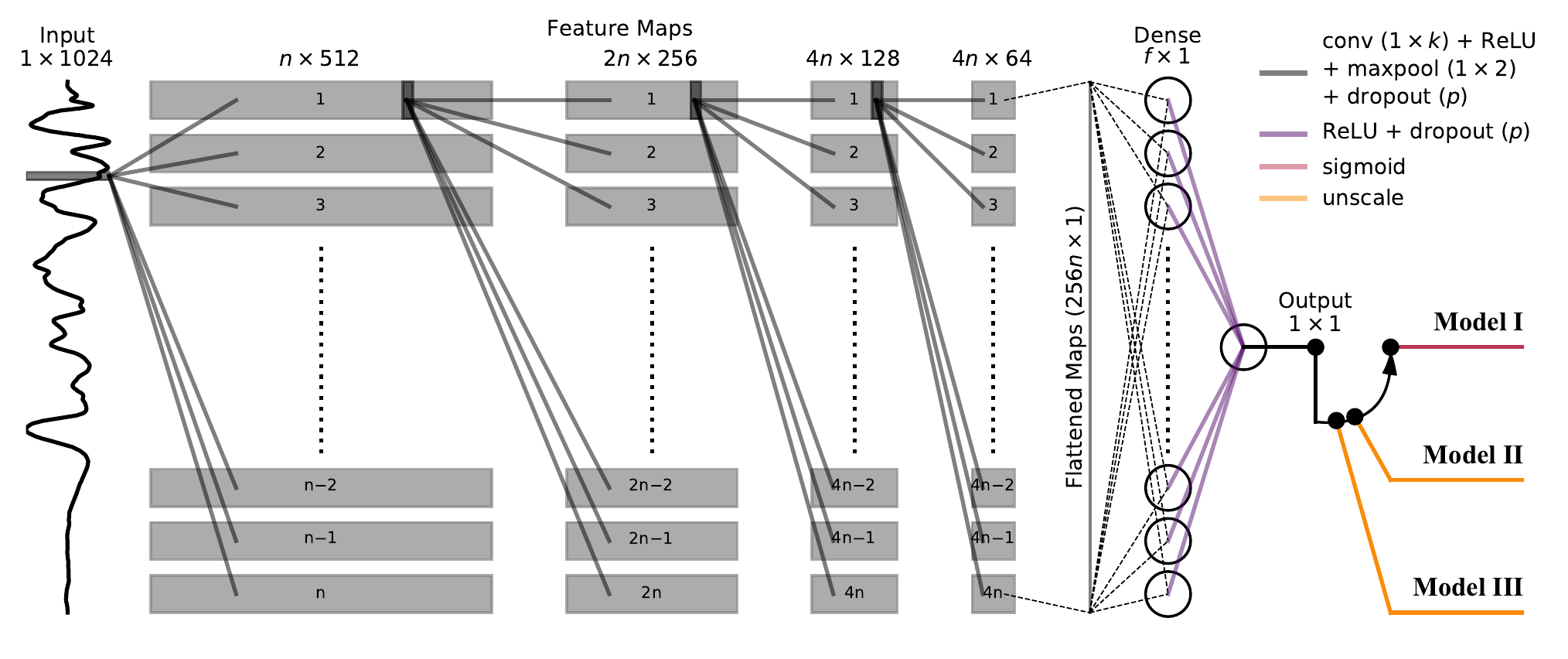}
 \caption{Schematic of the common neural network architecture used by each of the three fully independent networks that constitute Models~I--III (which collectively comprise {\tt deepSIP}). A set of $n$ feature maps is computed from the preprocessed input spectrum and down-sampled using max pooling (the resulting maps are indicated by grey boxes). This operation is then recursively applied to the down-sampled feature maps a total of three additional times, with the number of feature maps doubling for all but the last set. The final set of down-sampled feature maps is then flattened into a vector of length $256n$ and fed through a fully connected layer consisting of $f$ neurons before reaching the output neuron. Then, depending on the model, a final operation is performed to transform the raw output of the network into the appropriate context (``probabilities'' in the case of Model~I or dimensional phase or $\Delta m_{15}$ values for Models~II and III, respectively). The convolution kernel, $k$, number of feature maps generated by the first convolutional layer, $n$, and the number of neurons in the fully-connected layer, $f$, are all hyperparameters whose preferred value depends on the specific model. A dropout layer with dropout probability $p$ follows each weight layer aside from the output neuron.\label{fig:architecture}}
\end{figure*}

We present a schematic of our neural architecture, which utilises a total of four convolutional layers to extract representative features from input spectra, in Figure~\ref{fig:architecture}. We apply the Rectified Linear Unit \citep[ReLU,][]{ReLU} activation function to the output of each convolutional and fully connected layer. Following each convolutional layer we apply \hbox{max pooling} to reduce computational complexity and remove irrelevant features. Finally, we conclude each ``block'' (i.e., convolution + ReLU + max pool + dropout) in our network with a dropout layer to assist in the prevention of overfitting \citep{dropout}. Each convolutional and fully connected layer has its weights initialised with zero-centred Gaussian noise and its bias to a small, positive value. All of our models are implemented using {\tt PyTorch} 1.0 \citep{pytorch}, and we make our trained models and framework available as {\tt deepSIP}\footnote{\url{https://github.com/benstahl92/deepSIP}}, an open-source Python package (see Appendix~\ref{app:usage} for guidelines on basic usage).

Our selected architecture is largely motivated by insights gleaned from the sequences of variance spectra presented in Figure~\ref{fig:var-seq}. As noted earlier, the $\Delta m_{15}$-binned sequence shows more or less the same structure over our selected range of values. This homogeneity supports the use of a simple feed-forward network in the case of Model~II, but the depth of the network (i.e., how many convolutional layers are used) and the progression in the number of filters computed per layer are motivated by the \emph{heterogeneity} in features as they progress through the sequence. For example, the blue wing of the \ion{Si}{ii} $\uplambda 6355$ feature shows variation throughout the sequence, but the exact ``shape'' of that variation as a function of wavelength varies. For this reason, we use multiple convolutional layers and increase the number of convolutional kernels per layer in all but the last so that our networks have the capacity to make decisions based on many complex, highly-specialised features that are computed from a smaller number of basic features supplied by the earlier layers. The situation is mostly the same for Model~III, but the relevant variance spectra (i.e., those in the right column of Figure~\ref{fig:var-seq}) exhibit fewer common features and more extreme evolution in their shapes as the sequence progresses through phase bins. Motivated by this, we did carry out experiments with several architectures capable of predicting the phase in tandem with $\Delta m_{15}$, but none performed substantively better than our set of simple, independent networks. We do expect, however, that in addition to requiring much more high-quality training data, a specialised architecture would be crucial in expanding the output domain of Models~II and III. Indeed, extending the sequences of Figure~\ref{fig:var-seq} out to larger phase and $\Delta m_{15}$ values reveals significant feature evolution. A network capable of providing feedback between phase and $\Delta m_{15}$ predictions would allow for this evolution to be properly modeled.

The aforementioned dropout layers --- each of which randomly drops elements from their input with Bernoulli-distributed probability, $p$, and (in our implementation) rescales outputs by a factor of $1/(1 - p)$ during training --- serve a secondary purpose in our networks. Namely, this purpose is to make the networks probabilistic \citep{DVI}: each forward pass \emph{with dropout enabled in training mode} produces a different prediction, and thus, it is straightforward to quantitatively describe not only a point estimate (the prediction of a model) but also, an estimate of its uncertainty. To do so when generating predictions, we make $N$ stochastic (i.e., with dropout turned on) forward passes\footnote{We use a fiducial value of $N = 75$ when evaluating on the validation sets to select preferred \emph{training} hyperparameters (see Section~\ref{ssec:hyperparameter-selection}), but then treat it is a parameter to be further optimised prior to production-scale use (see Section~\ref{sssec:m23opthyperparams}).} for a given input and assign the mean and standard deviation of the resulting collection of predictions as the final model prediction and an estimate of its uncertainty.

\subsection{Training}
\label{ssec:training}

To train each of our models, we supply the appropriate training set in small batches and utilise an adaptive gradient descent algorithm \citep[ADAM;][]{ADAM} to minimise the appropriate objective function by updating the weights and biases in each layer of the network. For Model~I we employ the binary cross-entropy loss as our objective function, while for Models~II and III we use the mean squared error (MSE) loss. We also scale training outputs that are continuous (i.e., phase and $\Delta m_{15}$) such that they range from 0 to 1 using a transformation of the form $\mathbf{y}^\prime = (\mathbf{y} - y_{\rm min}) / (y_{\rm max} - y_{\rm min})$, where $(y_{\rm min}, y_{\rm max})$ represent the domain boundary (as shown in Figure~\ref{fig:dm15-phase-distr}) along the output's dimension. Subsequent predictions by these models are then unscaled using the inverse of this transformation. Model I predictions are transformed into ``probabilities'' using the sigmoid function.

We train each of our models for a total of 75 epochs, with the learning rate set to step down by a multiplicative factor of 0.1 after thresholds of 45, 60, and 70 epochs are reached. In testing, we found these choices to yield stable convergences without requiring excessive training time. At the culmination of each epoch we compute success metrics against the relevant validation set, thereby affording a specific measure of model-performance evolution in terms of the metrics we care most about (e.g., in dimensional, unscaled units for Models~II and III). For Model~I, we primarily use the area under the curve (AUC) of the Receiver Operating Characteristic (ROC) curve\footnote{An ROC curve shows the true-positive rate (ordinate) versus the false-positive rate (see Figure~\ref{fig:m1-roc}). The most optimal AUC score is 1, corresponding to a false-positive rate of 0 and a true positive rate of 1.}, whereas for Models~II and III, we primarily use the root-mean-square error (RMSE). We emphasise that although these are the primary metrics, we consider secondary indicators as well (see Sections~\ref{ssec:m1} \&~\ref{ssec:m23} for Models~I and II/III, respectively).

\subsection{Hyperparameter Selection}
\label{ssec:hyperparameter-selection}

There are several external parameters (i.e., not determined through backpropagation; henceforth referred to as ``hyperparameters'') that must be selected specifically for each of our models. Some are architectural (e.g., $k$, the size of each convolutional kernel; $n$, the number of distinct feature maps computed by the first convolutional layer; $f$, the number of neurons in the fully-connected layer; and $p$, the dropout probability used for training; see Figure~\ref{fig:architecture}), and some have to do with our training algorithm (e.g., training batch size, learning rate, and weight decay). The optimal choice of such parameters is not known {\it a priori} and is application dependent (e.g., we find that Model~II performs best when $k$ is less than the value that maximises Model~III performance; see Table~\ref{tab:hyperparams}). We note that the dropout probability used during training \emph{need not} be the same as that used when generating predictions. We therefore consider them separately as follows: the dropout probability for training is chosen as part of our hyperparameter selection process and the dropout probability used for generating predictions is separately chosen in tandem with $N$ (see Section~\ref{sssec:m23opthyperparams}). 

Thus, for each of Models~I, II, \& III, we perform a randomised search whereby we select preferred hyperparameter values by training and validating the models on many combinations of hyperparameters that are randomly drawn from a grid. A total of 12\,hr of compute time on a single of NVIDIA Tesla K80 GPU was allocated, per model, for these searches. We increase efficiency by automatically stopping training after 20 epochs when a performance threshold is not achieved on the validation set, and as a result, we are able to explore a significant portion of the selected hyperparameter space. Table~\ref{tab:hyperparams} details the full hyperparameter grid, and summarises the final set for each model. We discuss our selection criteria for determining these final, preferred sets in Sections~\ref{ssec:m1} \&~\ref{ssec:m23} for Models~I and II/III, respectively.

\begin{table}
\caption{Hyperparameter Grid.\label{tab:hyperparams}}
\begin{tabular}{ll}
\hline
\hline
Hyperparameter & Values$^a$\\
\hline
convolution kernel$^b$ ($k$) & $5,15,25^{{\rm I,II}},35^{{\rm III}}$ \\
filters in first convolution ($n$) & $8^{{\rm I,II,III}},16,32$ \\
fully connected neurons ($f$) & $16^{{\rm I,II}},32,64,128^{{\rm III}}$ \\
training dropout probability ($p$) & $0.01^{{\rm I,II,III}},0.05,0.1$ \\
batch size$^c$ & $2,4,8,16^{{\rm I,II}},32^{{\rm III}}$ \\
learning rate & $0.0005^{{\rm II,III}},0.001^{{\rm I}}$ \\
weight decay & $0.00001^{{\rm III}},0.0001^{{\rm I,II}}$ \\
\hline
\multicolumn{2}{p{8cm}}{$^a$Superscripts mark the preferred hyperparameter of the denoted model.}\\
\multicolumn{2}{p{8cm}}{$^b$Though small (e.g., $3\times3$) kernels are typical in 2D scenarios, significantly larger kernels have proven optimal in some 1D applications to astrophysical signals \citep[e.g., quasar spectra,][]{deepQuasar}.}\\
\multicolumn{2}{p{8cm}}{$^c$\citet{smallbatch} have suggested that batch sizes between 2 and 32 yield the best performance.}
\end{tabular}
\end{table}

\section{Results}
\label{sec:results}

\subsection{Model I: Domain Classification}
\label{ssec:m1}

\subsubsection{Preferred Hyperparameters}
\label{sssec:m1opthyperparams}

\begin{figure}
\centering
\includegraphics[width=\columnwidth]{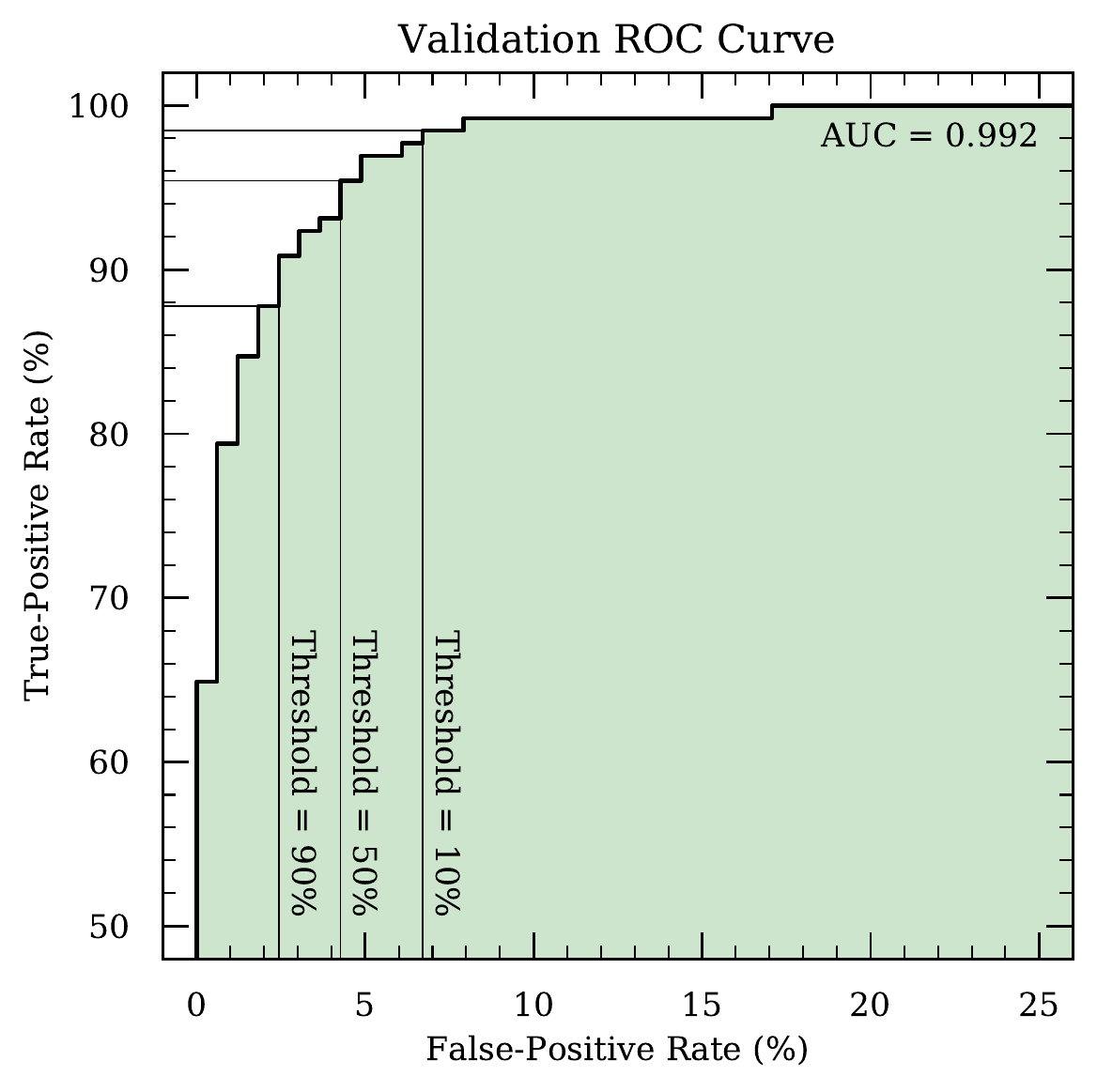}
 \caption{ROC curve for Model~I deployed on its validation set. The locations that correspond to various decision thresholds are indicated.\label{fig:m1-roc}}
\end{figure}

Our first model is designed to determine if an input spectrum belongs to a SN~Ia having $-10 \leq {\rm phase} < 18$\,d and $0.85 \leq \Delta m_{15} < 1.55$\,mag. If a spectrum satisfies these criteria, it is said to be \emph{in} the domain of interest; otherwise it is \emph{out}. In this way, Model~I serves as a precursor to Models~II and III: the subsequent predictions for spectra that it classifies as being outside the domain of interest should be carefully scrutinised, if not disregarded altogether. From our Model~I hyperparameter search (as discussed in Section~\ref{ssec:hyperparameter-selection} and summarised in Table~\ref{tab:hyperparams}), we find that the highest achieved validation ROC AUC score is 0.992 (see Figure~\ref{fig:m1-roc}), and on the basis of this score, we select the hyperparameters that yield it as the final, preferred set. These hyperparameters produce a network with a relatively modest number ($\sim 75$k) of trainable parameters. Though we currently do not consider uncertainty estimates for Model~I, we still use the full machinery of our probabilistic architecture with settings aligned to those of Models~II and III (i.e., $N = 30$ and $p = 0.02$; see Section~\ref{sssec:m23opthyperparams}) for consistency and to grant straightforward extensibility in the future.

\subsubsection{Decision Threshold}

With the information afforded by Figure~\ref{fig:m1-roc}, we are also able to tune the decision threshold of our model (i.e., the minimum ``probability'' of being \emph{in} to be classified as such). While many opt to use a default threshold of 50\% without further consideration, the optimal choice depends on striking an application-specific balance between the extent to which false positives can be tolerated and true positives can be missed. Taking a holistic view and recalling the aforementioned role of Model~I in the overall output of {\tt deepSIP}, it becomes obvious that the quality of Model~II and III predictions should be given the utmost priority. The optimal decision threshold in our case is therefore the one that yields the best Model~II and III performance on spectra that Model~I classifies as in-domain (which will be a mixture of true positives and false positives). This criterion is much more important than the overall classification accuracy\footnote{The accuracy score is the fraction of all predicted labels that are correct.} given the ``blurry'' nature of the domain boundary --- individual spectra can \emph{and do} fall so close to it that the particular side they end up on is determined by statistical variations.

To identify the optimal threshold, then, we use the Model~I validation\footnote{As discussed in Section~\ref{sssec:tvt}, our careful preparation of the Model~I validation and testing sets ensures that they are supersets of the corresponding sets for Models~II and III. Because of this, we can make predictions with the latter models on the former sets without concern for contamination.} set to study, as a function of decision threshold, Model~II/III RMSE scores segmented, based on the classifications of Model~I, into false negatives (FN), false positives (FP), true positives (TP), and all positives (P = FP + TP). We find that the Model~II scores belonging to FN and FP both follow a trend of decrease with rising decision threshold, and that those for FN are generally larger by a modest ($\sim 0.5$\,d) amount. The corresponding FN scores for Model~III also follow a decreasing trend (with rising threshold) but are significantly lower than those for the FP, which remain roughly constant at $\sim 0.2$\,mag until a dip forms between $\sim 90$--95\%. While one could argue that the FP scores for Model~II are acceptable over a wide range of thresholds ($\sim 2$--2.5\,d for thresholds of $\sim 5$--80\%, and lower thereafter), the aforementioned Model~III scores over a similar range are prohibitive. We simply cannot tolerate any significant contamination by FP with such high RMSE scores. This constrains the range of acceptable decision thresholds to just $\sim 90$--95\%, even at the expense of more FN with reasonable Model~II/III performance.

As the P scores are fairly flat between these bounds, we err to the low side (thereby minimising the number of incorrect classifications) and set our Model~I decision threshold to 90\%. This yields 13 FN (with RMSE scores of 1.91\,d and 0.080\,mag from Models~II and III, respectively) and 122 P (with scores of 1.05\,d and 0.068\,mag, respectively) of which 4 are FP (with scores of 1.39\,d and 0.160\,mag, respectively) from the 295 spectra (including 160 true negatives) in our validation set. Though this results in an accuracy score ($94.2\%$) that is slightly suboptimal to the peak value of 95.9\% achieved at a different threshold, it is still vastly in excess of the baseline score yielded by picking the most popular class every time ($55.6\%$) and it gives us confidence that the positives Model~I passes on to Models~II and III are sufficiently ``pure.''

\subsubsection{Performance on Testing Set}

With the decision threshold determined, we now make predictions on the testing set (which, as outlined in Section~\ref{sssec:tvt}, has \emph{not} been used to optimise the network or hyperparameters) and assess the efficacy of Model~I by comparing predicted labels to true labels. We find a similarly high ROC AUC of 0.989 and note that Model~I achieves a TP rate of 90.8\% at a FP rate of 2.4\% for our selected threshold of 90\%. From the 295 spectra in the testing set, Model~I delivers 160 true negatives, 12 FN, 4 FP, and 119 TP, collectively yielding an accuracy score of 94.6\% (as compared to the 55.6\% baseline score obtained by picking the most popular class every time). Of those marked as in-domain (i.e., P), Models~II and III yield RMSE scores of 1.06\,d and 0.072\,mag, respectively, while the TP subset performs even better at 1.00\,d and 0.064\,mag. These measures give us a high level of satisfaction with Model~I, and we therefore consider it complete.

\subsection{Models II \& III: Photometric Quantity Estimation}
\label{ssec:m23}

\subsubsection{Preferred Hyperparameters}
\label{sssec:m23opthyperparams}

Models~II and III are intended to determine the rest-frame phase and $\Delta m_{15}$, respectively, of a SN~Ia from its spectrum, assuming that it is within the phase and light-curve shape bounds that Model~I identifies (i.e., the spectrum is \emph{in} the relevant domain). As previously stated, our primary metric for regression tasks is the RMSE, but we consider two secondary indicators when selecting the final hyperparameter values: (i) the slope of a linear fit to predictions as a function of ground truth values, and (ii) the maximum absolute difference between predictions and labels (henceforth MR, for maximum residual). The first diagnoses the directionality of prediction errors --- systematic overestimates for low values and underestimates for high values are conveyed in a fitted slope of less than unity (and vice versa, though our models only bias in the aforementioned direction; see Section~\ref{sssec:biases}), and the second gives an indication of how homogeneous the absolute residuals are (when compared with the corresponding RMSE score). It is not sufficient for a set of hyperparameters to yield a competitive RMSE score; they must yield competitive scores across each of these three metrics.

We therefore identify the preferred hyperparameters for Models~II and III using a tiered approach to our search results. First, we filter to select only those results that have a slope above and an MR below a fiducial value when evaluated against the relevant validation set. Then, from the resulting subset, we select the entry with the lowest RMSE score. In this way, the final, preferred Model~II hyperparameters are chosen for yielding an RMSE of 1.15\,d, a slope of 0.96, and an MR of 4.32\,d and the Model~III hyperparameters on the basis of yielding scores of 0.065\,mag, 0.823, and 0.206\,mag, respectively. The final networks have $\sim 75$k and $\sim 320$k trainable parameters, respectively. Although it is beyond the scope of this study to make any definitive or in-depth statements about the significance of the final hyperparameters, it is interesting to note the differences between those that yield the best observed performance in Models~II and III. For example, we find that Model~III performs best with a larger convolution kernel, $k$, than does Model~II (35 for Model~III versus 25 for Model~II); this may indicate that features encoding phase information are generally narrower than those which encode $\Delta m_{15}$. At the same time, Model~II requires fewer neurons in the fully connected layer than does Model~III. This is consistent with our general intuition that phases are more ``simply'' codified in spectral features than $\Delta m_{15}$ (or other luminosity indicators), which may be best parameterised by ratios of nonlocal features \citep[as suggested by][]{Nugent1995}.

\begin{figure}
\centering
\includegraphics[width=\columnwidth]{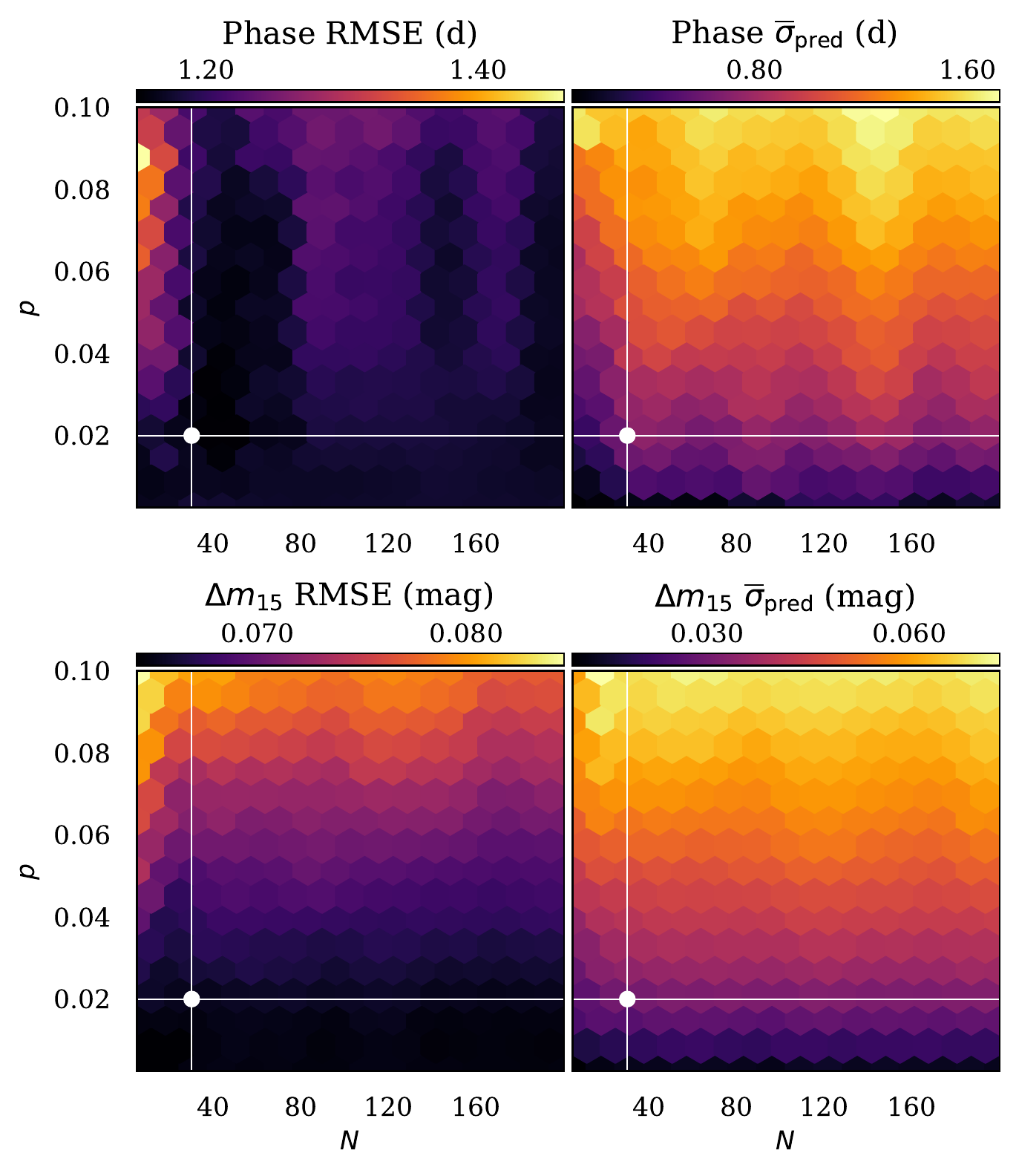}
 \caption{Validation RMSE and mean predicted uncertainty values for Models~II and III over a grid of $N$ and $p$ values. A separate colour bar is provided for each panel, and the selected $(N, p)$ set is indicated.\label{fig:mcnum-p-sweep}}
\end{figure}

As a final refinement to the parameters that govern Models~II and III, we study the effect of varying the number of stochastic forward passes ($N$) and dropout probability ($p$) when the models are used to generate predictions. To do so, we use Models~II and III to generate predictions from the relevant validation set over a grid of $(N, p)$ values and then tabulate the RMSE and mean estimated uncertainty at each point. The results (visualised in Figure~\ref{fig:mcnum-p-sweep}) are generally consistent with our expectations: mean predicted uncertainties steadily grow with $p$ as do RMSE values, albeit at a much less significant rate. For Model~III, both metrics show minimal dependence on $N$, but for Model~II there are ``bands'' of improved RMSE performance at $N = 30$--80, 140--160, and beyond 180 (though they only outperform their surroundings by $\lesssim 0.2$\,d). While our primary concern is optimising model performance (i.e., achieving low RMSE scores and ``reasonable'' uncertainty estimates), it is desirable from a compute-time perspective to use the lowest $N$ value possible. We therefore select $(N, p) = (30, 0.02)$ for all {\tt deepSIP} predictions. As Figure~\ref{fig:mcnum-p-sweep} clearly shows, this yields the desired low $N$ (for fast prediction times) without compromising the quality of model predictions.

\subsubsection{Performance on Testing Set}

\begin{figure*}
\centering
\includegraphics[width=\textwidth]{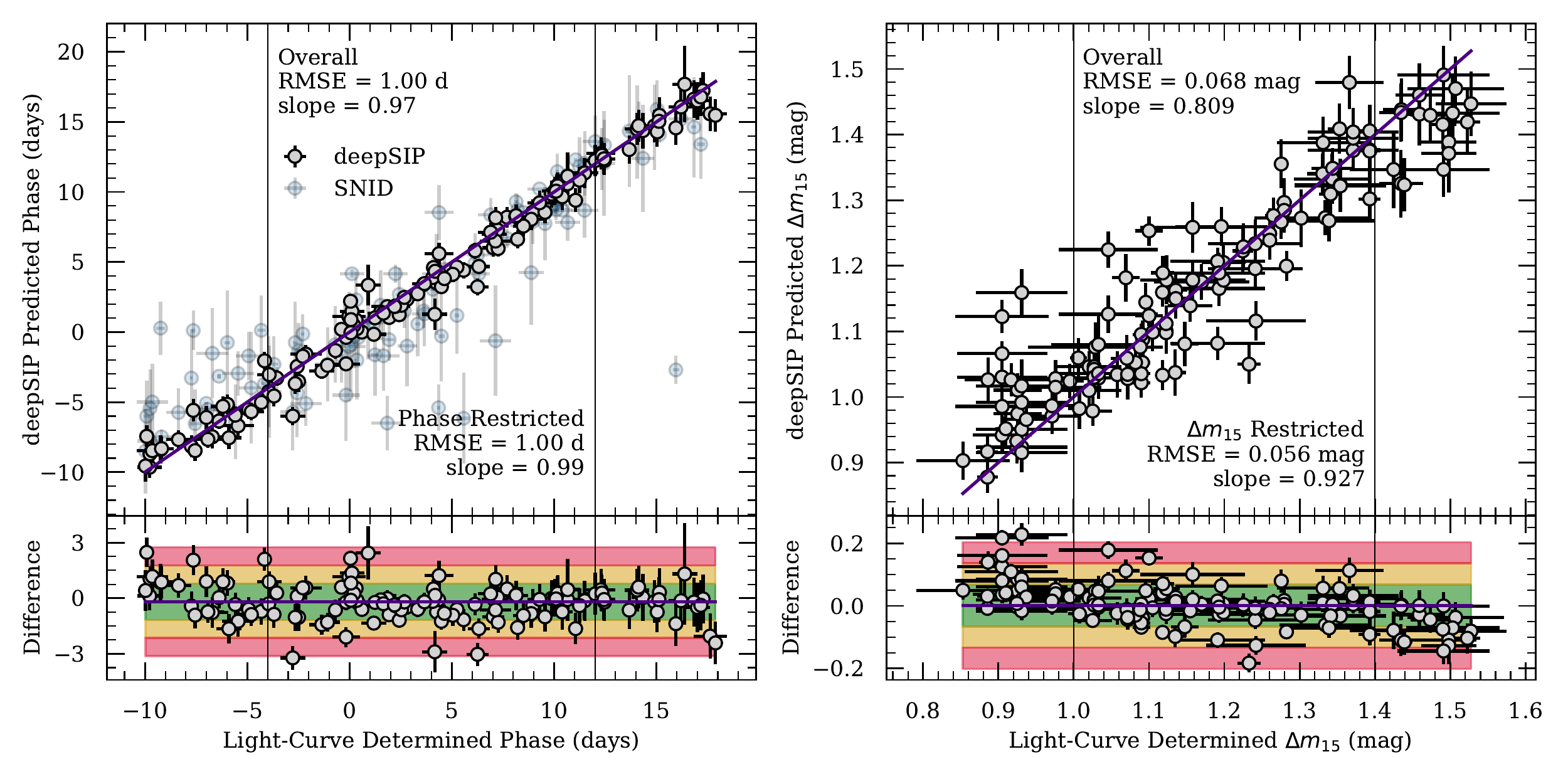}
 \caption{Phase (left panel) and $\Delta m_{15}$ (right panel) determined by {\tt deepSIP} versus ground truth values from the respective Model~II and III testing sets, with residuals in the lower panels. The indigo line in each upper plot shows the one-to-one correspondence for ground truth values, while in the lower plots, it indicates the median residual. The green, yellow, and red regions indicate the $1\sigma, 2\sigma$, and $3\sigma$ bounds about the median residual, respectively. We mark the relevant RSME and slope, both globally and for a restricted subset (see Section~\ref{sssec:biases}) in each panel. {\tt SNID}-based phase predictions are presented as an overlay to the upper-left panel, but they are omitted from the residuals for clarity.\label{fig:phase-dm15-comp}}
\end{figure*}

Having selected the final hyperparameters for each model, we make predictions on the relevant testing sets. As shown in Figure~\ref{fig:phase-dm15-comp}, we find strong agreement when we compare predictions to ground truth labels, achieving RMSE, slope, and MR scores of 1.00\,d, 0.97, and 3.22\,d; and 0.068\,mag, 0.809, and 0.228\,mag, respectively. For the 131 samples in each testing set, the median phase ($\Delta m_{15}$) residual is $-0.19$\,d ($0.001$\,mag); also, 94 (94) are within one standard deviation of the median, 122 (123) are within two, and 130 (129) are within three.

In Section~\ref{sssec:pred-unc}, we investigate the quality of the uncertainty estimates produced by our models, which serve as a systematic error probe by quantifying the dispersion between $N$ realisations drawn (for each input spectrum) from the underlying distribution that \emph{is} each of our models. Here we do the opposite in an attempt to quantify statistical error: we remove the stochasticity from our models (by disabling dropout) and assess prediction robustness when they are fed perturbed inputs. We generate such inputs using our data-augmentation strategy (see Section~\ref{sssec:augmentation}) to bootstrap our Model~II and III testing sets up to 5000 instances, each of which is slightly perturbed in redshift, noise, and signal length. The results, which we assess by means of the RMSE, are highly satisfactory: Model~II yields 1.05\,d and Model~III delivers 0.080\,mag, both broadly consistent with the corresponding measures reported above. One could potentially use a variation of this strategy to generate a unique statistical uncertainty estimate for each input spectrum and include that with each prediction, but we defer that task to future study and development.

\subsubsection{Comparison with {\tt SNID}-derived Phases}

To contextualise the level of performance of Model~II with regard to phase predictions, we attempt to characterise the spectra in the testing set using a series of {\tt SNID} runs that adhere to the specifications\footnote{The {\tt SNID} procedure is minimally intrusive, but intended to increase reliability by determining the type, subtype, redshift, and phase from a SN~Ia spectrum in consecutive runs which progressively refine the set of templates used for comparison.} laid out by S20. When we do so, we find that {\tt deepSIP} performs \emph{significantly} better than {\tt SNID} in virtually every way. Whereas the {\tt SNID}-based scheme (consisting of a total of four runs per spectrum, all of which are controlled and read using a {\tt Python} script and hence subject to small {\tt Python} overheads) takes $\sim 7$\,min to process all 131 testing samples on a server with a modern CPU, our Model~II (which entails a total of 30 stochastic forward passes per spectrum) takes under 1\,s on a single NVIDIA Tesla K80 GPU to characterise preprocessed spectra ($<1$\,min on a modern, four-core CPU), with $< 1$\,s for preprocessing. Moreover, the {\tt SNID} approach fails to derive a phase in 18/131 instances while {\tt deepSIP} is successful in making a prediction in all cases. Perhaps most significantly, the RMSE between those instances where {\tt SNID} successfully predicts a phase and the true phases is $3.48$\,d, a factor of $\sim 3.5$ times worse (in RMSE; $\sim 12$ times worse in MSE) than that for our {\tt deepSIP}-derived results. The {\tt SNID} results are also much more afflicted by a bias to overestimate the earliest phases (a tendency that has been observed and discussed by S20 and others); the mean residual (predicted minus true) at phases from $-10$\,d to $-5$\,d is $+3.30$\,d for {\tt SNID}-derived results, but just $+0.31$\,d for those from {\tt deepSIP}. This bias is also exhibited in the aforementioned fitted-slope metric, with {\tt SNID} yielding a value of 0.80 compared to 0.97 for Model~II.

\subsubsection{$\Delta m_{15}$ Consistency}

Unfortunately, we are unable to perform an analogous {\tt SNID}-based comparison for $\Delta m_{15}$ values, but we can exploit a unique feature that the $\Delta m_{15}$ labels possess to perform a separate test on the validity of Model~III. Unlike phases which are are unique to individual spectra, $\Delta m_{15}$ values are unique to individual SNe~Ia. As a result, it is not uncommon for multiple spectra in the training, validation, or testing sets to map to the same $\Delta m_{15}$ value. We can test how well Model~III deals with this degeneracy by looking at the scatter (parameterised by the standard deviation) in $\Delta m_{15}$ predictions in the testing set, grouped by distinct $\Delta m_{15}$ label (and therefore, by distinct object).

When we do so, we find encouraging results which we summarise with the following observations: (i) as compared to the global scatter in predicted $\Delta m_{15}$ values (0.161\,mag), the median scatter in predicted values \emph{per} distinct $\Delta m_{15}$ label is just 0.018\,mag; (ii) the observed distribution is positively skewed so that the majority of the scatter is near zero (e.g., the 25th percentile is 0.007\,mag while the 75th percentile is 0.045\,mag); and (iii) of the two examples with scatter $> 0.07$\,mag, both are at or near the extremes of the predicted $\Delta m_{15}$ values (i.e., where the model's predictions are typically the most uncertain and the training data are sparsest).

\subsubsection{Biases}
\label{sssec:biases}

Though mitigated by our selection criteria (namely, our choice to enforce a 40 spectrum-per-bin saturation policy), Model~II and (especially) Model~III do exhibit some bias toward the more central values in their prediction ranges. This is unfortunate, but expected given the nonuniformity of our training data (e.g., see Figures~\ref{fig:dm15-phase-distr} \&~\ref{fig:unc-vs-prediction}). We emphasise that despite this bias, the \emph{residual} distributions are approximately symmetric. Still, it is useful to quantify the extent of this bias, and we choose to do so by means of the previously mentioned slope of a linear fit to the results presented in Figure~\ref{fig:phase-dm15-comp}. 

In doing so, we find a slope of 0.97 for phases and 0.809 for $\Delta m_{15}$ values, confirming our suspicion that the bias is present (and more pronounced in Model~III). However, if we select a more restrictive subset to exclude the biased ends (taken to be the equivalent of 1.5 bins from Figure~\ref{fig:dm15-phase-distr}; i.e., 6\,d and 0.15\,mag from each end), the fitted slopes improve to 0.99 and 0.927, respectively. The Model~III RMSE value improves as well, dropping to 0.056\,mag. Users of {\tt deepSIP} may therefore choose to give more weight to results within these restricted ranges. Such improvements reinforce our belief that more performance could be extracted from our models with a larger and more balanced training set.

\subsubsection{Estimated Uncertainties}
\label{sssec:pred-unc}

The aforementioned metrics are quite satisfactory, but we must also verify the quality of the uncertainty estimates produced by our models. A basic measure of this is afforded by comparing RMSE to \emph{weighted} RMSE values (henceforth, wRMSE), defined by
\begin{equation}
	{\rm wRMSE} = \sqrt{\frac{\sum_{i = 1}^N (\hat{y}_i - y_i)^2 \hat{\sigma}^{-2}_i}{\sum_{i = 1}^N \hat{\sigma}^{-2}_i}},
\end{equation}
where $\mathbf{y}$ are ground truth labels, and $(\mathbf{\hat{y}}, \hat{\sigma})$ are the corresponding predicted labels and estimated uncertainties (i.e., the mean and standard deviation of the stochastic samples generated as described in Section~\ref{ssec:architecture}). A situation where wRMSE $>$ RMSE would reflect poorly on the uncertainty estimates because it would imply, in aggregate, an inverse correlation between them and residuals (i.e., model predictions are generally \emph{more} wrong where the model is \emph{more} certain); conversely, a situation where wRMSE $<$ RMSE is an affirmation (but not conclusive determination) of the quality of our uncertainty estimates because it suggests that model predictions are generally \emph{more} correct where the model is \emph{more} certain. In our case, the results are favorable (albeit at only a modest level): Model~II yields a wRMSE score of 0.92\,d while Model~III yields 0.065\,mag.

\begin{figure*}
\centering
\includegraphics[width=\textwidth]{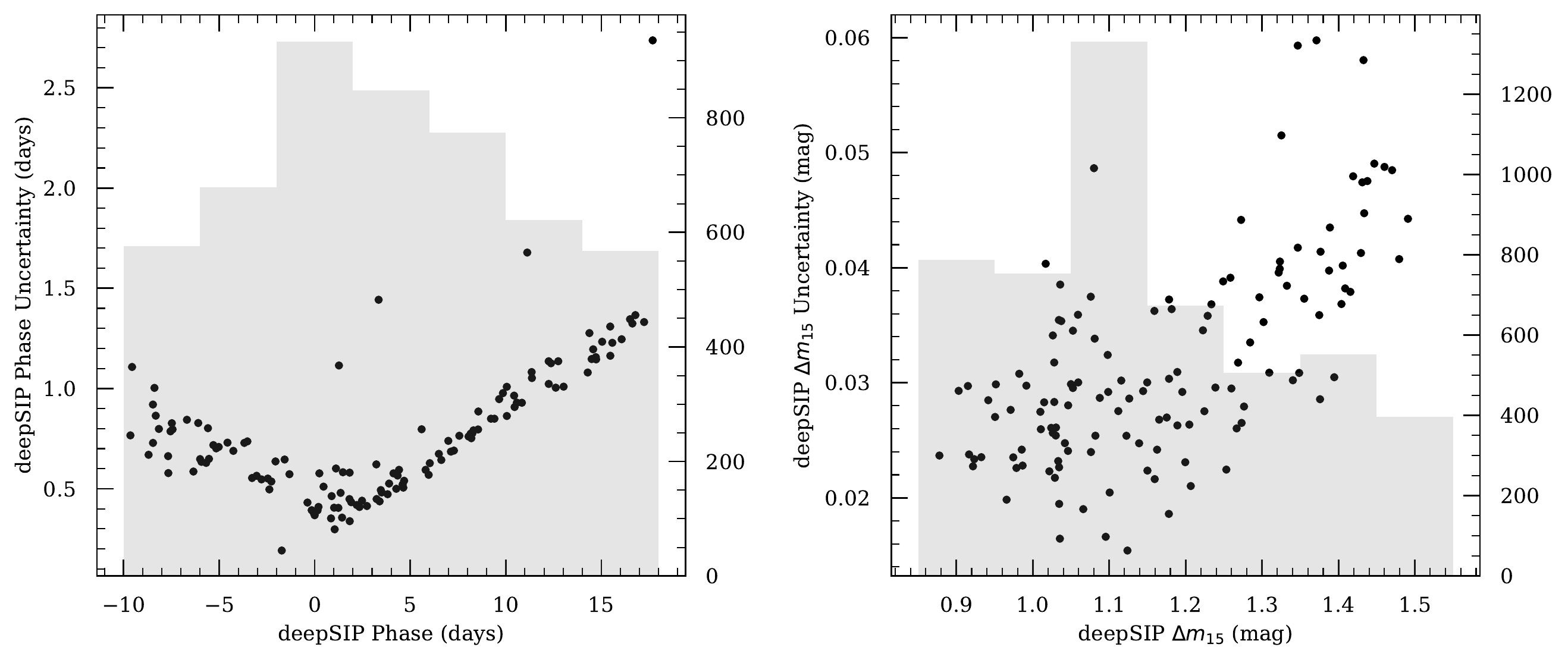}
 \caption{{\tt deepSIP}-determined phase (left panel) and $\Delta m_{15}$ (right panel) uncertainties versus predicted values from the respective Model~II and III testing sets. The grey distributions behind each convey the relevant training data (including those from the augmented training set).\label{fig:unc-vs-prediction}}
\end{figure*}

To further probe the quality of our estimated uncertainties, one might be tempted to study the relationship between estimated and true uncertainties (i.e., those derived from light-curve fits). This, however, would not be appropriate because the true uncertainties were not accounted for by our loss function during model training (nor anywhere else aside from selection cuts). Instead, we can ask a more appropriate question of our estimated uncertainties: \emph{how do they behave relative to the data our models were trained on?} The answer, depicted in Figure~\ref{fig:unc-vs-prediction}, is encouraging. We see that our estimated uncertainties for both models are generally smallest in the region where training data are most abundant, and that the uncertainties grow steadily as the training data become more scarce. This captures the general behaviour we desire, though the uncertainties may be modestly underestimated (given the lower panels in Figure~\ref{fig:phase-dm15-comp}, but we defer an extensive study of this to future work). Thus, on the basis of this desired behaviour and all prior points elucidated above, we consider Models~II and III ready for deployment.

\section{Conclusion}
\label{sec:conclusion}

In this paper we present and characterise the performance of {\tt deepSIP}, an open-source software package that encapsulates a set of three CNNs that collectively map optical SN~Ia spectra to their corresponding phases and light-curve shapes (parameterised by $\Delta m_{15}$). The treatment of these tasks with supervised learning --- and the specific use of a CNN architecture --- is a natural choice in many regards. This choice is rewarded with highly satisfactory performance.

To train, validate, and test our models, we compile a significant collection of low-redshift SN~Ia spectra by drawing from public data releases from the CfA, CSP, and our own BSNIP. These spectra form, after preprocessing, the inputs of the input-output pairs that our models learn to map. To assemble the corresponding outputs (i.e., phases and light-curve shapes), we draw from the photometric data releases of the same research campaigns, supplementing with five SNe~Ia for which we publish light curves that have recently become available in our own archives. We fit all light curves (except for those from S19 and CSP3 who performed the same fits) using the {\tt SNooPy} $E(B - V)$ model, ensuring systematic consistency between all phase and $\Delta m_{15}$ values used herein. After all cuts are accounted for, our final compilation consists of 2754 spectra with photometrically-derived phase and $\Delta m_{15}$ values, and of these, 1113 are within the phase and $\Delta m_{15}$ constrained domain of interest for Models~II and III.

Because we draw spectra from multiple sources, we take great care to both understand and mitigate systematic differences between sources so that our models form decision paths exclusively from physically significant features encoded in the spectra. Our mitigation strategy manifests chiefly in our preprocessing procedure which, among other things, discards any spectral information below 3450\,\AA\, or above 7500\,\AA. Though painful, this prevents our models from being affected by the presence or absence of signal at more extreme wavelengths --- a distinction which on the red end almost perfectly segments CfA, CSP, and BSNIP spectra. We also let our desire for telescope agnosticism guide the data augmentation strategy that we employ to increase our modest sample size. In addition to a redshift perturbation, we vary the extent to which augmented spectra are smoothed and randomly drop signal from the ends. The latter two actions serve to blur out the signature imparted by the specific equipment used to collect a given spectrum.

We describe the common neural network architecture which underlies each of our models and whose organisation and layout is largely motivated by our consideration of sequences of variance spectra. Under the assumption that spectral variations in SNe~Ia are \emph{mostly} a result of phase and light-curve shape (i.e., luminosity, via a width-luminosity relation), such sequences inform us about which regions in spectra most strongly encode a given target. Our observation of a relatively consistent set of features in such sequences has informed our adoption of a rather simple feed-forward network. To estimate uncertainties alongside point values for our targets, we use a probabilistic model provided by dropout variational inference. We employ a randomised grid search to determine the preferred set of hyperparameters for each of Models I--III, and upon doing so, set out to assess their performance.

To do so, we deploy each model against a distinct (i.e., unused during training and hyperparameter selection) testing set. In the case of Model~I (a binary classifier of ``in/out'' with regard to a domain defined by $-10 \leq {\rm phase} < 18$\,d and $0.85 \leq \Delta m_{15} < 1.55$\,mag), we achieve an accuracy score of 94.6\% and ROC AUC of 0.989. At a false-positive rate of 2.4\%, Model~I has an in-domain detection rate of 90.8\%. With Model~II (a continuous predictor of phases from $-10$\,d to 18\,d), we achieve an RMSE (wRMSE) of 1.00\,d (0.92\,d), a marked improvement over {\tt SNID}-derived predictions on the same spectra. Finally, for Model~III (a continuous predictor of $\Delta m_{15}$ values from 0.85\,mag to 1.55\,mag), we achieve an RMSE (wRMSE) of 0.068\,mag (0.065\,mag). These final, trained models are publicly available through {\tt deepSIP} which provides an easy-to-use API for deploying them to characterise new SN~Ia spectra. We strongly encourage public use of {\tt deepSIP} for this purpose.

Looking to the future, we expect that the performance of {\tt deepSIP} could be significantly improved as more spectra with corresponding light curves become available. Indeed, the dominant factor in our selection of the phase--$\Delta m_{15}$ domain inside of which Models~II and III offer predictions is the paucity of data available with more extreme light-curve shapes. As such data become more prevalent, the networks which underly {\tt deepSIP} can easily be retrained and if necessary, modified to accommodate feedback between predicted phase and $\Delta m_{15}$ values that may be necessary given the substantial feature evolution observed beyond the domain boundary. We welcome community involvement on these fronts (accumulating more data and designing more sophisticated network architectures), and intend to continue active, transparent development on our publicly hosted {\tt GitHub} repository.

\section*{Acknowledgements}

We thank our referee, Tom Charnock, whose exceptionally insightful comments and suggestions improved the manuscript. B.E.S. thanks Marc J. Staley for generously providing fellowship funding and D. Muthukrishna for helpful discussions about CNN architecture choices. J.M.P., J.S.B. were partially supported by a Gordon and Betty Moore Foundation Data-Driven Discovery grant. Support for A.V.F.'s supernova group has been provided by U.S. National Science Foundation
(NSF) grant AST--1211916, the Richard and Rhoda Goldman Fund, the TABASGO Foundation, Gary and Cynthia
Bengier (T.deJ. is a Bengier Postdoctoral Fellow), the Christopher R.
Redlich Fund, and the Miller Institute for Basic Research in Science
(U.C. Berkeley). KAIT and its              
ongoing operation were made possible by donations from Sun                    
Microsystems, Inc., the Hewlett-Packard Company, AutoScope                    
Corporation, Lick Observatory, the NSF, the University of California,         
the Sylvia \& Jim Katzman Foundation, and the TABASGO Foundation. 
Research at Lick Observatory is partially supported by a generous           
gift from Google.
In addition, we greatly appreciate contributions from
numerous individuals, including
Charles Baxter and Jinee Tao,
George and Sharon Bensch, 
Firmin Berta,
Marc and Cristina Bensadoun,
Frank and Roberta Bliss,
Eliza Brown and Hal Candee,
Kathy Burck and Gilbert Montoya,
Alan and Jane Chew,
David and Linda Cornfield,
Michael Danylchuk,
Jim and Hildy DeFrisco,
William and Phyllis Draper,
Luke Ellis and Laura Sawczuk,
Jim Erbs and Shan Atkins,
Alan Eustace and Kathy Kwan,
Peter and Robin Frazier, 
David Friedberg,
Harvey Glasser,
Charles and Gretchen Gooding,
Alan Gould and Diane Tokugawa,
Thomas and Dana Grogan,
Timothy and Judi Hachman, 
Alan and Gladys Hoefer,
Charles and Patricia Hunt,
Stephen and Catherine Imbler,
Adam and Rita Kablanian,
Roger and Jody Lawler,
Kenneth and Gloria Levy,
Peter Maier,
DuBose and Nancy Montgomery,
Rand Morimoto and Ana Henderson,
Sunil Nagaraj and Mary Katherine Stimmler,
Peter and Kristan Norvig,
James and Marie O'Brient,
Emilie and Doug Ogden,
Paul and Sandra Otellini,
Jeanne and Sanford Robertson,
Sissy Sailors and Red Conger, 
Stanley and Miriam Schiffman,
Thomas and Alison Schneider,
Ajay Shah and Lata Krishnan,
Alex and Irina Shubat,
the Silicon Valley Community Foundation,
Mary-Lou Smulders and Nicholas Hodson,
Hans Spiller,
Alan and Janet Stanford,
the Hugh Stuart Center Charitable Trust,
Clark and Sharon Winslow,
Weldon and Ruth Wood,
David and Angie Yancey, 
and many others.

We thank
Stanley Browne,
Sanyum Channa,
Ian Crossfield,
Edward Falcon,
Tatiana Gibson,
Ellen Glad,
Christopher Griffith,
Julia Hestenes,
Benjamin Jeffers,
Charles Kilpatrick,
Michelle Kislak,
Laura Langland,
Joel Leja,
Gary Li,
Michael Ross,
Timothy Ross,
Costas Soler,
Samantha Stegman,
Kevin Tang,
Patrick Thrasher,
Priscilla West,
Sameen Yunus,
and Keto Zhang,
for their effort in obtaining data with the 1\,m Nickel telescope at
Lick Observatory.

This research used the Savio computational cluster resource provided by the Berkeley Research Computing program at U.C. Berkeley (supported by the U.C. Berkeley Chancellor, Vice Chancellor for Research, and Chief Information Officer). This research has made use of the CfA Supernova Archive, which is funded in part by the NSF through grant AST--0907903. The Pan-STARRS1 Surveys (PS1) and the PS1 public science archive have been made possible through contributions by the Institute for Astronomy, the University of Hawaii, the Pan-STARRS Project Office, the Max-Planck Society and its participating institutes, the Max Planck Institute for Astronomy, Heidelberg and the Max Planck Institute for Extraterrestrial Physics, Garching, the Johns Hopkins University, Durham University, the University of Edinburgh, Queen's University Belfast, the Harvard-Smithsonian Center for Astrophysics, the Las Cumbres Observatory Global Telescope Network Incorporated, the National Central University of Taiwan, the Space Telescope Science Institute, the National Aeronautics and Space Administration (NASA) under grant NNX08AR22G issued through the Planetary Science Division of the NASA Science Mission Directorate, NSF grant AST--1238877, the University of Maryland, E\"otv\"os Lorand University (ELTE), the Los Alamos National Laboratory, and the Gordon and Betty Moore Foundation.

\section*{Data Availability}

The data published herein are available in the article and in its online supplementary material.




\bibliographystyle{mnras}
\bibliography{ML_bib}




\appendix

\section{Supplementary Light Curves}
\label{app:newlc}

In the time since the photometric dataset presented by S19 was published, we have continued to obtain host-galaxy template images for the unpublished SNe~Ia in our archives. With these new observations, we are able to process five additional SNe~Ia (SN~2007S, SN~2008hv, SN~2010kg, SN~2017hpa, and SN~2018oh.) of utility to this work. We therefore present their \emph{BVRI} light curves (some also have unfiltered observations, which we refer to as {\it Clear}). All processing steps are identical to those described by S19, so we provide only a brief summary of the methodology before delivering results.

All images were collected using either the 0.76\,m Katzman Automatic Imaging Telescope \citep[KAIT;][]{Li-KAIT,Filippenko-KAIT} or the 1\,m Nickel telescope, both of which are located at Lick Observatory where the seeing averages $\sim 2^{\prime\prime}$. After removing bias and dark current, flat-field correcting, and deriving an astrometric solution, we pass images to our automated photometry pipeline\footnote{\url{https://github.com/benstahl92/LOSSPhotPypeline}} ({\tt LOSSPhotPypeline}; see S19), which handles all aspects of the remaining processing.

With host-galaxy template images obtained on dark nights using the Nickel telescope, the pipeline removes contaminating flux due to a SN's host galaxy and then performs point-spread function (PSF) photometry using procedures from the IDL Astronomy User's Library\footnote{\url{https://idlastro.gsfc.nasa.gov/homepage.html}} to measure the SN's flux relative to selected standard stars in the same field. The resulting instrumental magnitudes are calibrated with at least two (but often more) of the selected standard stars from the Pan-STARRS1 Survey \citep[PS1;][]{PS1}. To do this, PS1 magnitudes are transformed to the \citet{Landolt1992} system using the prescription given by \citet{Tonry2012}, and then into the appropriate natural-system magnitudes using coupled equations of the form
\begin{align}
  b &= B + C_B(B - V) + {\rm constant},\\
  v &= V + C_V(B - V) + {\rm constant},\\
  r &= R + C_R(V - R) + {\rm constant, and}\\
  i &= I + C_I(V - I) + {\rm constant},
\end{align}
where natural (Landolt) system magnitudes are expressed in lower (upper)-case letters and $C_X$ is the linear colour term corresponding to filter $X$ (given in S19's Table 1). Temporally close ($< 0.4$\,d) observations in the \emph{same} passband are averaged together and then those in \emph{distinct} passbands are grouped by their midpoint epoch to form natural-system light curves. Finally, the aforementioned equations are inverted to yield standardised light curves on the Landolt system. The uncertainties on each magnitude in our light curves are ultimately derived from three sources: ``statistical'' (e.g., Poisson variations in observed brightness, scatter in sky values, uncertainty in sky brightness), ``calibration'' (e.g., derived colour terms, uncertainty in PS1 magnitudes), and ``simulation'' (as described by S19).

\begin{figure*}
 \includegraphics[width=\textwidth]{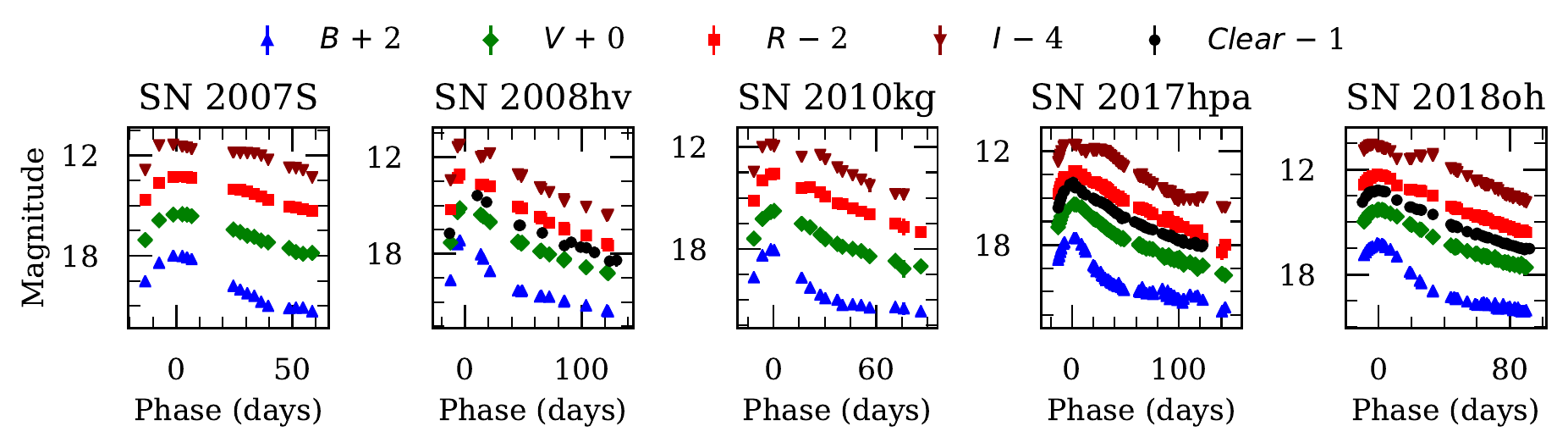}
 \caption{Observed light curves for the previously unpublished SNe~Ia in our archives. In most cases the error bars are smaller than the points themselves. All dates have been shifted relative to the time of maximum \emph{B}-band brightness. The light curves are available in tabular form through our SNDB and are also accessible as online supplementary material.\label{fig:newlc}}
\end{figure*}

We present the final Landolt-system light curves derived from the aforementioned processing steps in Figure~\ref{fig:newlc}. The final light curves are publicly available through our U.C. Berkeley SuperNova DataBase\footnote{\url{http://heracles.astro.berkeley.edu/sndb/}} \citep[SNDB; S12;][]{Shivvers2016SNDB} and in the Supplementary Materials included with this article. We describe our method for (and results from) fitting these light curves (and the others in our compilation) using {\tt SNooPy} in the following section.

\section{Light-Curve Fitting}
\label{app:snoopy}

We use the $E(B - V)$ model as implemented by the {\tt SNooPy} package to simultaneously fit the \emph{BVRI} (or subset thereof) light curves in our photometry compilation. The model assumes a peak \emph{B}-band magnitude and $B - X$ colours parameterised by the decline rate, and the results from fitting are the time of maximum \emph{B}-band light ($t_{\rm max}$), decline-rate parameter\footnote{Though already stated, we emphasise again that $\Delta m_{15}$ may deviate from $\Delta m_{15}(B)$, as discussed by \citet{SNooPy} and subsequently verified by S19.} ($\Delta m_{15}$), host-galaxy reddening, and distance modulus. We use the fitting results obtained by S19 and CSP for their datasets, and employ the strategy of the former to fit the remaining SNe~Ia in our photometry compilation (i.e., those from G10 and CfA1-3). We give the resulting values of $t_{\rm max}$ and $\Delta m_{15}$ for all SNe which pass a visual inspection for fit quality in Table~\ref{tab:snoopy-fits}, and defer a more thorough explanation of the fitting process to S19 (and their listed references).

\onecolumn
\begin{landscape}
{\footnotesize
\begin{longtable}{lccc|lccc|lccc}
\caption{{\tt SNooPy} Fitted Parameters.\label{tab:snoopy-fits}}\\
\hline
\hline
\multicolumn{1}{l}{SN} & \multicolumn{1}{c}{Source$^a$} & \multicolumn{1}{c}{$t_{\rm max}$ (MJD)} & \multicolumn{1}{r}{$\Delta m_{15}$ (mag)} & \multicolumn{1}{l}{SN} & \multicolumn{1}{c}{Source$^a$} & \multicolumn{1}{c}{$t_{\rm max}$ (MJD)} & \multicolumn{1}{r}{$\Delta m_{15}$ (mag)} & \multicolumn{1}{l}{SN} & \multicolumn{1}{c}{Source$^a$} & \multicolumn{1}{c}{$t_{\rm max}$ (MJD)} & \multicolumn{1}{r}{$\Delta m_{15}$ (mag)}\\
\hline
\endfirsthead
\hline
\hline
\multicolumn{1}{l}{SN} & \multicolumn{1}{c}{Source$^a$} & \multicolumn{1}{c}{$t_{\rm max}$ (MJD)} & \multicolumn{1}{r}{$\Delta m_{15}$ (mag)} & \multicolumn{1}{l}{SN} & \multicolumn{1}{c}{Source$^a$} & \multicolumn{1}{c}{$t_{\rm max}$ (MJD)} & \multicolumn{1}{r}{$\Delta m_{15}$ (mag)} & \multicolumn{1}{l}{SN} & \multicolumn{1}{c}{Source$^a$} & \multicolumn{1}{c}{$t_{\rm max}$ (MJD)} & \multicolumn{1}{r}{$\Delta m_{15}$ (mag)}\\
\hline
\endhead
\hline
\multicolumn{12}{c}{Table \thetable~continued}
\endfoot
\hline
\multicolumn{12}{p{22cm}}{Note: only fitted parameters used in our final compilation are presented. See CSP3 and S19 for the corresponding fits for their datasets.}\\
\multicolumn{12}{p{22cm}}{$^a$Sources of light curves used for fitting. Those marked by ``this'' refer to those that we publish here (see Appendix~\ref{app:newlc}).}\\
\endlastfoot
 1993ac &  CfA1 &  $49269.34 \pm 0.70$ &  $1.098 \pm 0.040$ &  2001bg &   G10 &  $52039.95 \pm 0.24$ &  $1.155 \pm 0.029$ &           2004dt &   G10 &  $53240.19 \pm 0.18$ &  $1.136 \pm 0.022$ \\
 1993ae &  CfA1 &  $49289.15 \pm 0.19$ &  $1.480 \pm 0.022$ &  2001br &   G10 &  $52051.78 \pm 0.16$ &  $1.435 \pm 0.024$ &           2004fu &  CfA3 &  $53325.28 \pm 0.48$ &  $1.170 \pm 0.063$ \\
  1994M &  CfA1 &  $49474.56 \pm 0.29$ &  $1.331 \pm 0.021$ &  2001cj &   G10 &  $52064.98 \pm 0.11$ &  $0.903 \pm 0.013$ &           2004fz &   G10 &  $53333.53 \pm 0.10$ &  $1.473 \pm 0.027$ \\
  1994Q &  CfA1 &  $49495.35 \pm 0.40$ &  $1.093 \pm 0.020$ &  2001ck &   G10 &  $52072.10 \pm 0.12$ &  $1.079 \pm 0.017$ &           2005bc &   G10 &  $53469.96 \pm 0.12$ &  $1.646 \pm 0.028$ \\
  1994S &  CfA1 &  $49517.90 \pm 0.36$ &  $1.021 \pm 0.038$ &  2001cp &   G10 &  $52087.82 \pm 0.08$ &  $0.884 \pm 0.009$ &           2005bo &   G10 &  $53478.10 \pm 0.16$ &  $1.270 \pm 0.008$ \\
 1994ae &  CfA1 &  $49685.19 \pm 0.10$ &  $1.058 \pm 0.011$ &  2001da &   G10 &  $52107.02 \pm 0.19$ &  $1.149 \pm 0.021$ &           2005cf &   G10 &  $53533.85 \pm 0.08$ &  $1.123 \pm 0.008$ \\
  1995D &  CfA1 &  $49767.78 \pm 0.16$ &  $0.886 \pm 0.012$ &  2001dl &   G10 &  $52130.58 \pm 0.12$ &  $1.022 \pm 0.023$ &           2005de &   G10 &  $53598.65 \pm 0.10$ &  $1.216 \pm 0.012$ \\
  1995E &  CfA1 &  $49774.49 \pm 0.22$ &  $1.069 \pm 0.018$ &  2001eh &   G10 &  $52168.42 \pm 0.16$ &  $0.837 \pm 0.006$ &           2005eu &   G10 &  $53659.81 \pm 0.12$ &  $1.099 \pm 0.018$ \\
 1995ac &  CfA1 &  $49992.44 \pm 0.30$ &  $0.941 \pm 0.036$ &  2001en &   G10 &  $52192.43 \pm 0.10$ &  $1.282 \pm 0.005$ &           2005hf &  CfA3 &  $53660.68 \pm 0.95$ &  $1.449 \pm 0.053$ \\
 1995ak &  CfA1 &  $50022.22 \pm 0.50$ &  $1.278 \pm 0.018$ &  2001ep &   G10 &  $52199.65 \pm 0.17$ &  $1.133 \pm 0.023$ &           2005ls &  CfA3 &  $53714.40 \pm 0.31$ &  $0.930 \pm 0.033$ \\
 1995al &  CfA1 &  $50028.26 \pm 0.19$ &  $0.910 \pm 0.014$ &  2001ex &   G10 &  $52204.27 \pm 0.28$ &  $1.813 \pm 0.029$ &           2005lz &  CfA3 &  $53735.82 \pm 0.39$ &  $1.276 \pm 0.031$ \\
 1995bd &  CfA1 &  $50086.33 \pm 0.15$ &  $0.937 \pm 0.030$ &  2001fe &  CfA3 &  $52229.01 \pm 0.30$ &  $0.956 \pm 0.019$ &           2005mc &  CfA3 &  $53733.83 \pm 0.22$ &  $1.733 \pm 0.026$ \\
  1996C &  CfA1 &  $50127.77 \pm 0.33$ &  $0.965 \pm 0.019$ &   2002G &   G10 &  $52297.43 \pm 0.43$ &  $1.145 \pm 0.050$ &           2005ms &  CfA3 &  $53744.16 \pm 0.10$ &  $1.079 \pm 0.018$ \\
  1996X &  CfA1 &  $50190.73 \pm 0.13$ &  $1.225 \pm 0.009$ &  2002aw &   G10 &  $52324.57 \pm 0.25$ &  $1.123 \pm 0.017$ &           2005mz &  CfA3 &  $53745.01 \pm 0.13$ &  $1.864 \pm 0.003$ \\
 1996ai &  CfA1 &  $50256.52 \pm 0.38$ &  $1.112 \pm 0.036$ &  2002bf &   G10 &  $52335.09 \pm 0.00$ &  $1.093 \pm 0.032$ &            2006X &   G10 &  $53786.01 \pm 0.55$ &  $0.971 \pm 0.038$ \\
 1996bk &  CfA1 &  $50369.07 \pm 0.55$ &  $1.758 \pm 0.010$ &  2002bo &   G10 &  $52356.29 \pm 0.12$ &  $1.105 \pm 0.014$ &           2006ac &   G10 &  $53779.74 \pm 0.51$ &  $1.199 \pm 0.029$ \\
 1996bl &  CfA1 &  $50376.23 \pm 0.19$ &  $1.100 \pm 0.019$ &  2002bz &  CfA3 &  $52368.19 \pm 0.53$ &  $1.366 \pm 0.045$ &           2006al &  CfA3 &  $53789.06 \pm 0.35$ &  $1.569 \pm 0.044$ \\
 1996bo &  CfA1 &  $50386.51 \pm 0.38$ &  $1.156 \pm 0.036$ &  2002cd &   G10 &  $52384.39 \pm 0.23$ &  $1.101 \pm 0.024$ &           2006az &  CfA3 &  $53826.76 \pm 0.13$ &  $1.354 \pm 0.027$ \\
 1996bv &  CfA1 &  $50403.42 \pm 0.39$ &  $0.930 \pm 0.023$ &  2002cf &   G10 &  $52384.39 \pm 0.10$ &  $1.823 \pm 0.001$ &           2006bb &  CfA3 &  $53815.83 \pm 0.48$ &  $1.615 \pm 0.018$ \\
 1997bp &  CfA2 &  $50550.08 \pm 0.43$ &  $1.114 \pm 0.049$ &  2002cr &   G10 &  $52409.07 \pm 0.09$ &  $1.260 \pm 0.007$ &           2006bt &   G10 &  $53857.71 \pm 0.23$ &  $1.091 \pm 0.036$ \\
 1997bq &  CfA2 &  $50558.43 \pm 0.31$ &  $1.136 \pm 0.031$ &  2002cs &   G10 &  $52410.26 \pm 0.17$ &  $1.097 \pm 0.020$ &           2006cc &  CfA3 &  $53874.13 \pm 0.13$ &  $1.044 \pm 0.030$ \\
 1997br &  CfA2 &  $50559.90 \pm 0.27$ &  $1.122 \pm 0.027$ &  2002cu &   G10 &  $52416.12 \pm 0.10$ &  $1.461 \pm 0.022$ &           2006cp &   G10 &  $53896.91 \pm 0.31$ &  $1.130 \pm 0.054$ \\
 1997cw &  CfA2 &  $50627.98 \pm 0.44$ &  $0.811 \pm 0.020$ &  2002de &   G10 &  $52432.99 \pm 0.16$ &  $1.071 \pm 0.021$ &           2006dm &   G10 &  $53928.20 \pm 0.09$ &  $1.523 \pm 0.017$ \\
 1997do &  CfA2 &  $50766.18 \pm 0.23$ &  $1.088 \pm 0.023$ &  2002dj &   G10 &  $52450.79 \pm 0.35$ &  $1.149 \pm 0.046$ &           2006ef &   G10 &  $53968.14 \pm 0.22$ &  $1.273 \pm 0.012$ \\
 1997dt &  CfA2 &  $50786.77 \pm 0.23$ &  $1.341 \pm 0.054$ &  2002dl &   G10 &  $52451.92 \pm 0.09$ &  $1.759 \pm 0.007$ &           2006ej &   G10 &  $53975.67 \pm 0.17$ &  $1.498 \pm 0.037$ \\
 1998ab &  CfA2 &  $50914.39 \pm 0.19$ &  $1.103 \pm 0.021$ &  2002do &   G10 &  $52441.42 \pm 0.47$ &  $1.718 \pm 0.010$ &           2006em &   G10 &  $53976.32 \pm 0.24$ &  $1.823 \pm 0.001$ \\
 1998bp &  CfA2 &  $50936.36 \pm 0.18$ &  $1.800 \pm 0.012$ &  2002dp &   G10 &  $52450.38 \pm 0.11$ &  $1.214 \pm 0.008$ &           2006en &   G10 &  $53970.97 \pm 0.34$ &  $0.974 \pm 0.021$ \\
 1998de &   G10 &  $51025.70 \pm 0.12$ &  $1.821 \pm 0.001$ &  2002eb &   G10 &  $52494.31 \pm 0.08$ &  $1.067 \pm 0.012$ &           2006gr &   G10 &  $54012.41 \pm 0.13$ &  $1.084 \pm 0.017$ \\
 1998dh &   G10 &  $51029.00 \pm 0.12$ &  $1.118 \pm 0.015$ &  2002ef &   G10 &  $52489.88 \pm 0.17$ &  $1.144 \pm 0.019$ &           2006hb &   G10 &  $54000.62 \pm 0.29$ &  $1.693 \pm 0.011$ \\
 1998dk &  CfA2 &  $51057.17 \pm 0.29$ &  $1.135 \pm 0.015$ &  2002el &   G10 &  $52507.93 \pm 0.07$ &  $1.367 \pm 0.020$ &           2006le &   G10 &  $54047.74 \pm 0.16$ &  $1.082 \pm 0.018$ \\
 1998dm &   G10 &  $51060.25 \pm 0.12$ &  $1.008 \pm 0.015$ &  2002er &   G10 &  $52524.49 \pm 0.16$ &  $1.140 \pm 0.018$ &           2006lf &   G10 &  $54045.03 \pm 0.17$ &  $1.459 \pm 0.032$ \\
 1998ec &   G10 &  $51088.65 \pm 0.93$ &  $1.146 \pm 0.072$ &  2002eu &   G10 &  $52520.22 \pm 0.24$ &  $1.731 \pm 0.010$ &           2006mo &  CfA3 &  $54048.02 \pm 0.35$ &  $1.653 \pm 0.047$ \\
 1998ef &   G10 &  $51113.19 \pm 0.10$ &  $1.280 \pm 0.007$ &  2002fb &   G10 &  $52529.02 \pm 0.09$ &  $1.824 \pm 0.000$ &           2006mp &  CfA3 &  $54053.92 \pm 0.12$ &  $0.995 \pm 0.019$ \\
 1998eg &   G10 &  $51110.13 \pm 0.70$ &  $1.117 \pm 0.047$ &  2002fk &   G10 &  $52547.13 \pm 0.10$ &  $1.027 \pm 0.010$ &           2006oa &  CfA3 &  $54066.52 \pm 0.19$ &  $0.953 \pm 0.054$ \\
 1998es &   G10 &  $51142.61 \pm 0.07$ &  $0.925 \pm 0.010$ &  2002ha &   G10 &  $52580.77 \pm 0.04$ &  $1.362 \pm 0.008$ &           2006qo &  CfA3 &  $54082.94 \pm 0.15$ &  $1.054 \pm 0.014$ \\
 1999aa &   G10 &  $51231.89 \pm 0.13$ &  $0.886 \pm 0.014$ &  2002he &   G10 &  $52585.40 \pm 0.06$ &  $1.439 \pm 0.011$ &           2006sr &  CfA3 &  $54092.38 \pm 0.14$ &  $1.279 \pm 0.011$ \\
 1999ac &   G10 &  $51249.98 \pm 0.19$ &  $1.104 \pm 0.022$ &  2002hu &  CfA3 &  $52592.12 \pm 0.13$ &  $1.089 \pm 0.015$ &           2006td &  CfA3 &  $54099.32 \pm 0.14$ &  $1.422 \pm 0.020$ \\
 1999by &   G10 &  $51307.95 \pm 0.10$ &  $1.824 \pm 0.000$ &  2002hw &  CfA3 &  $52595.63 \pm 0.12$ &  $1.552 \pm 0.027$ &           2006te &  CfA3 &  $54096.89 \pm 0.40$ &  $1.130 \pm 0.021$ \\
 1999cc &  CfA2 &  $51315.33 \pm 0.18$ &  $1.344 \pm 0.028$ &  2002jg &   G10 &  $52609.62 \pm 0.07$ &  $1.417 \pm 0.019$ &            2007O &   G10 &  $54122.77 \pm 0.40$ &  $1.139 \pm 0.041$ \\
 1999cl &   G10 &  $51340.88 \pm 0.22$ &  $1.144 \pm 0.026$ &  2002jy &  CfA3 &  $52634.02 \pm 0.37$ &  $0.881 \pm 0.026$ &            2007S &  this &  $54144.73 \pm 0.24$ &  $0.836 \pm 0.005$ \\
 1999cp &   G10 &  $51362.61 \pm 0.12$ &  $1.032 \pm 0.021$ &  2002kf &  CfA3 &  $52638.27 \pm 0.43$ &  $1.236 \pm 0.016$ &           2007al &  CfA3 &  $54169.59 \pm 0.25$ &  $1.857 \pm 0.016$ \\
 1999da &   G10 &  $51369.79 \pm 0.09$ &  $1.823 \pm 0.001$ &   2003W &   G10 &  $52679.65 \pm 0.23$ &  $1.130 \pm 0.029$ &           2007ap &  CfA3 &  $54168.31 \pm 0.15$ &  $1.490 \pm 0.018$ \\
 1999dg &   G10 &  $51392.64 \pm 0.32$ &  $1.509 \pm 0.049$ &   2003Y &   G10 &  $52676.54 \pm 0.12$ &  $1.822 \pm 0.001$ &           2007au &   G10 &  $54183.75 \pm 0.17$ &  $1.754 \pm 0.017$ \\
 1999dk &   G10 &  $51414.72 \pm 0.13$ &  $1.103 \pm 0.014$ &  2003cg &   G10 &  $52729.24 \pm 0.14$ &  $1.136 \pm 0.015$ &           2007bj &   G10 &  $54199.92 \pm 0.70$ &  $0.905 \pm 0.039$ \\
 1999dq &   G10 &  $51435.95 \pm 0.10$ &  $1.090 \pm 0.012$ &  2003ch &  CfA3 &  $52725.42 \pm 0.35$ &  $1.274 \pm 0.019$ &           2007bz &  CfA3 &  $54213.68 \pm 0.50$ &  $0.888 \pm 0.051$ \\
 1999ej &   G10 &  $51482.78 \pm 0.06$ &  $1.565 \pm 0.010$ &  2003cq &   G10 &  $52737.64 \pm 0.33$ &  $1.170 \pm 0.023$ &           2007ci &   G10 &  $54246.24 \pm 0.18$ &  $1.732 \pm 0.015$ \\
 1999ek &  CfA2 &  $51481.16 \pm 0.39$ &  $1.164 \pm 0.016$ &  2003du &   G10 &  $52765.71 \pm 0.11$ &  $1.055 \pm 0.007$ &           2007co &   G10 &  $54264.07 \pm 0.13$ &  $1.108 \pm 0.020$ \\
 1999gd &  CfA2 &  $51518.85 \pm 0.53$ &  $1.168 \pm 0.025$ &  2003fa &   G10 &  $52806.74 \pm 0.08$ &  $1.072 \pm 0.009$ &           2007cq &   G10 &  $54280.01 \pm 0.16$ &  $1.148 \pm 0.018$ \\
 1999gh &  CfA2 &  $51513.29 \pm 0.37$ &  $1.737 \pm 0.008$ &  2003gn &   G10 &  $52852.44 \pm 0.09$ &  $1.399 \pm 0.028$ &           2007fr &   G10 &  $54301.70 \pm 0.16$ &  $1.755 \pm 0.013$ \\
 1999gp &   G10 &  $51549.68 \pm 0.12$ &  $1.076 \pm 0.014$ &  2003gs &   G10 &  $52847.65 \pm 0.25$ &  $1.820 \pm 0.001$ &           2007qe &   G10 &  $54429.37 \pm 0.24$ &  $1.128 \pm 0.029$ \\
 2000ce &  CfA2 &  $51666.53 \pm 0.46$ &  $1.015 \pm 0.034$ &  2003gt &   G10 &  $52861.61 \pm 0.06$ &  $1.095 \pm 0.008$ &           2007sr &   G10 &  $54447.66 \pm 0.33$ &  $1.085 \pm 0.015$ \\
 2000cf &  CfA2 &  $51671.86 \pm 0.29$ &  $1.144 \pm 0.022$ &  2003he &   G10 &  $52875.89 \pm 0.11$ &  $0.956 \pm 0.016$ &            2008C &   G10 &  $54463.84 \pm 0.53$ &  $1.100 \pm 0.019$ \\
 2000cn &   G10 &  $51706.76 \pm 0.08$ &  $1.713 \pm 0.008$ &  2003hv &   G10 &  $52890.04 \pm 0.09$ &  $1.554 \pm 0.008$ &            2008L &   G10 &  $54493.79 \pm 0.14$ &  $1.545 \pm 0.023$ \\
 2000cp &   G10 &  $51719.52 \pm 0.68$ &  $1.158 \pm 0.070$ &  2003ic &  CfA3 &  $52906.73 \pm 0.71$ &  $1.425 \pm 0.058$ &            2008Q &   G10 &  $54504.62 \pm 0.23$ &  $1.029 \pm 0.090$ \\
 2000cu &   G10 &  $51743.78 \pm 0.10$ &  $1.502 \pm 0.017$ &  2003it &  CfA3 &  $52934.97 \pm 0.21$ &  $1.435 \pm 0.029$ &            2008Z &   G10 &  $54514.74 \pm 0.21$ &  $1.020 \pm 0.043$ \\
 2000cw &   G10 &  $51747.89 \pm 0.17$ &  $1.153 \pm 0.022$ &  2003iv &  CfA3 &  $52933.94 \pm 0.20$ &  $1.527 \pm 0.047$ &           2008af &  CfA3 &  $54503.68 \pm 0.73$ &  $1.532 \pm 0.041$ \\
 2000cx &   G10 &  $51752.60 \pm 0.15$ &  $1.265 \pm 0.011$ &  2003kf &   G10 &  $52980.13 \pm 0.22$ &  $1.025 \pm 0.025$ &           2008ar &   G10 &  $54534.35 \pm 0.17$ &  $1.113 \pm 0.023$ \\
 2000dk &   G10 &  $51811.79 \pm 0.05$ &  $1.712 \pm 0.005$ &   2004E &   G10 &  $53014.95 \pm 0.43$ &  $1.121 \pm 0.023$ &           2008dr &   G10 &  $54649.53 \pm 0.22$ &  $1.463 \pm 0.035$ \\
 2000dm &   G10 &  $51815.25 \pm 0.11$ &  $1.535 \pm 0.017$ &   2004S &   G10 &  $53039.57 \pm 0.26$ &  $1.115 \pm 0.013$ &           2008ec &   G10 &  $54673.78 \pm 0.09$ &  $1.341 \pm 0.016$ \\
 2000dn &   G10 &  $51824.52 \pm 0.15$ &  $1.107 \pm 0.023$ &  2004as &   G10 &  $53084.66 \pm 0.18$ &  $1.133 \pm 0.037$ &           2008ei &   G10 &  $54670.78 \pm 0.75$ &  $1.142 \pm 0.058$ \\
 2000dr &   G10 &  $51833.97 \pm 0.00$ &  $1.753 \pm 0.007$ &  2004at &   G10 &  $53091.65 \pm 0.07$ &  $1.092 \pm 0.009$ &           2008hv &  this &  $54817.01 \pm 0.08$ &  $1.276 \pm 0.007$ \\
 2000fa &   G10 &  $51891.78 \pm 0.12$ &  $0.974 \pm 0.012$ &  2004bd &   G10 &  $53096.93 \pm 0.39$ &  $1.736 \pm 0.007$ &           2010kg &  this &  $55543.89 \pm 0.19$ &  $1.269 \pm 0.014$ \\
  2001E &   G10 &  $51926.16 \pm 0.41$ &  $1.021 \pm 0.058$ &  2004bg &   G10 &  $53108.35 \pm 0.22$ &  $1.024 \pm 0.018$ &          2017hpa &  this &  $58066.66 \pm 0.05$ &  $1.101 \pm 0.005$ \\
  2001V &   G10 &  $51971.43 \pm 0.30$ &  $0.849 \pm 0.013$ &  2004bk &   G10 &  $53111.45 \pm 0.46$ &  $0.892 \pm 0.019$ &           2018oh &  this &  $58163.16 \pm 0.07$ &  $1.064 \pm 0.008$ \\
 2001ah &   G10 &  $52005.19 \pm 0.21$ &  $0.921 \pm 0.036$ &  2004br &   G10 &  $53147.60 \pm 0.23$ &  $0.880 \pm 0.022$ &  SNF20071021-000 &   G10 &  $54406.73 \pm 0.18$ &  $1.180 \pm 0.017$ \\
 2001az &  CfA3 &  $52031.82 \pm 0.59$ &  $1.016 \pm 0.043$ &  2004bv &   G10 &  $53160.48 \pm 0.10$ &  $1.083 \pm 0.010$ &  SNF20080514-002 &   G10 &  $54611.84 \pm 0.13$ &  $1.393 \pm 0.015$ \\
 2001bf &   G10 &  $52044.66 \pm 0.00$ &  $0.921 \pm 0.025$ &  2004bw &   G10 &  $53162.66 \pm 0.11$ &  $1.323 \pm 0.012$ &  SNF20080909-030 &   G10 &  $54730.08 \pm 0.62$ &  $0.926 \pm 0.032$ \\
\end{longtable}}
\end{landscape}
\twocolumn

\section{Usage}
\label{app:usage}

In tandem with this paper, we provide a well-documented\footnote{\url{https://deepsip.readthedocs.io}} and easy-to-use {\tt Python} package called {\tt deepSIP}. The final, trained models presented herein are shipped with the code base, and hence, it is ready for deployment on new spectra. To use {\tt deepSIP} for this purpose, one must prepare {\tt spectra} as a {\tt pandas} DataFrame with three mandatory columns: {\tt SN, filename, z} for the for name(s) of the SN(e)~Ia, their filenames, and their redshifts (it is assumed that the spectra need to be de-redshifted), respectively. Generating predictions is then accomplished as follows:

\begin{lstlisting}[language=Python, numbers=none]
from deepSIP import deepSIP
ds = deepSIP()
predictions = ds.predict(spectra, status=True)
\end{lstlisting}

All necessary spectral preprocessing steps are performed automatically prior to generating predictions. No arguments are necessary to instantiate {\tt deepSIP} under normal use cases (though one may give the keyword argument {\tt drop\_rate} to change the dropout probability). When generating predictions from {\tt spectra}, three keyword arguments can --- but need not be --- invoked: (i) {\tt threshold} sets the decision threshold for Model~I (0.9 by default), (ii) {\tt mcnum} sets the number of stochastic forward passes (30 by default), and (iii) {\tt status} can be used to enable status bars. The returned {\tt predictions} are provided as a {\tt pandas} DataFrame with five columns: {\tt Domain, prob\_Domain, Phase, e\_Phase, dm15, e\_dm15}, corresponding to the respective predictions of Models~I--III. Each row in {\tt predictions} corresponds to the spectrum from the same row in {\tt spectra}.


\bsp	
\label{lastpage}
\end{document}